\documentclass[final]{siamltex}
\usepackage{graphicx}
\usepackage{amsmath,amssymb,euscript,yfonts,psfrag,dsfont,graphicx,bbm,color,amstext,wasysym,pdfsync}
\usepackage[usenames,dvipsnames]{xcolor}
\usepackage{longtable}
\usepackage{makecell}
\usepackage{slashbox,booktabs}
\usepackage{amsfonts,amsmath,mathbbol,mathtools}
\usepackage{subfig}
\usepackage{balance}

\newcommand{\xb}{\mathbf  x}

\newcommand{\Xc}{\mathcal X}
\newcommand{\Yc}{\mathcal Y}

\newcommand{\Sc}{\mathcal S}
\newcommand{\Wc}{\mathcal W}

\newcommand{\Mc}{\mathcal M}

\newcommand{\mOne}{{\mathds{1}}}

\newcommand{\cX}{{\mathcal X}}
\newcommand{\cY}{{\mathcal Y}}
\newcommand{\cZ}{{\mathcal Z}}
\newcommand{\Ebb}{{\mathbb E}\,}

\newcommand{\bmat}{\left[ \begin{matrix}}
\newcommand{\emat}{\end{matrix} \right]}
\newcommand{\trace}{\operatorname{trace}}
\newcommand{\card}{{\operatorname{card}}}

\title{Convex Clustering via Optimal Mass Transport}
\author{Francesca P. Carli, Lipeng Ning, and Tryphon T. Georgiou
\thanks{Department of Electrical \& Computer Eng., University of Minnesota, Minneapolis, MN 55455;
\hspace*{15pt}Email: \{fpcarli, ningx015, tryphon\}@umn.edu}; 
Supported in part by the NSF and the AFOSR.}

\begin{document}
\maketitle
\thispagestyle{empty}
\pagestyle{empty}

\begin{abstract}
We consider approximating distributions within the framework of optimal mass transport and specialize to the problem of clustering data sets. 
Distances between distributions are measured in the Wasserstein metric. 
The main problem we consider is that of approximating sample distributions by ones with sparse support. This provides a new viewpoint to clustering. We propose different relaxations of a cardinality function which penalizes the size of the support set. We establish that a certain relaxation provides the tightest convex lower approximation to the cardinality penalty.
We compare the performance of alternative relaxations on a numerical study on clustering. 
\end{abstract}

\section{Introduction}

The analysis of data sets invariably requires approximating observed sample distributions by ones that belong to a particular family. Instances include modeling using sums of Gaussians. While such families can in principle be quite general, the metric used to quantify mismatch ought to reflect appropriate features. A natural geometry is that provided by optimal mass transport endowing the space of distribution with, for instance, the Wasserstein metric $d_{\Wc_2}$. Throughout we will use this metric to study approximation problems. We specialize to a family $\Sc$ of distributions with sparse support on discrete spaces with application to clustering.

In more detail, we consider the problem of approximating a given sample distribution $p_0$  with a distribution $p_1$ belonging to a class $\Sc$ by solving the problem
\begin{equation}\label{probl:OMT_Wass_dist_with_constraints}
\underset{p_1 \in \Sc \subset \mathcal{P}}{\min} \;\,\, d_{\Wc_2}\left(p_0,p_1\right).
\end{equation}
Throughout, our spaces are discrete and $\mathcal{P}$ denotes the probability simplex.  
Membership in $\Sc$ can typically be relaxed by introducing a suitable penalty function 
\begin{equation}\label{probl:OMT_penalized_Wass_dist}
\underset{p_1 \in \mathcal{P}}{\min} \;\,\, d_{\Wc_2}\left(p_0,p_1\right) + \lambda \mathcal{I}(p_1)
\end{equation}
where the function $\mathcal{I}$ penalizes $p_1\notin\Sc$ and $\lambda>0$. 
In this paper, we specialize to the case where the set $\Sc$ represents the family of distributions having a sparse support on a discrete space, which leads to a cardinality penalized optimization problem. 
The sparse atoms of the support correspond to the representatives of different clusters. 
Association to those clusters' representatives is provided by the solution to the aforementioned optimal mass transportation problem and is dictated by the optimal transportation plan (see below for details). 

A heuristic for cardinality that has attracted a lot of attention in recent years is $\ell_1$--norm regularization \cite{Tibshirani1996,Chen1998,Candes2005,Bruckstein2009,Chandrasekaran2010}. 
However, this cannot be used to promote sparsity on a probability simplex since the $\ell_1$ norm of a probability measure is always one (see e.g., the recent paper by Pilanci et al. \cite{ElGhaoui2012NIPS}).
In the present paper, we propose relaxations of the cardinality penalty on the probability simplex. 
The main idea is to express the cardinality penalty in terms of the optimal transportation plan $\Pi$ between the sample distribution and a target distribution with required sparsity properties. The transportation plan $\Pi$ is itself a probability distribution on a larger space. 
We show that \emph{convex} relaxations of the cardinality penalty can  be realized via a 
sum--of--norms penalty on the transportation plan matrix and via the introduction of a suitable indicator function. 
This leads to a \emph{convex} optimization scheme to solve the clustering problem.   
We show that the sum--of--norms relaxation provides the \emph{tightest} convex lower approximation of the original cardinality penalty.
Finally, we present numerical examples that underscore the effectiveness and relevance of the proposed approaches with regard to the problem of clustering data sets. 

The paper is organized as follows. 
In Section \ref{sec:preliminaries} we introduce the optimal mass transportation problem. 
Optimal transport with a cardinality penalty is introduced in Section \ref{sec:cardinality_penalized_probl_and_relax}
and different relaxations are proposed. 
In particular a relaxation based on a sum--of--norms penalty is discussed in Section \ref{sec:GLasso_relaxation} 
while a relaxation based on the introduction of auxiliary boolean variables is discussed in Section \ref{sec:y_relaxation}. 
In Section \ref{sec:l_inf_relaxation} we adapt the approach in \cite{ElGhaoui2012NIPS} to our setting and compare with the proposed techniques.  
In Section \ref{sec:clustering_via_OMT} we specialize to the clustering problem of and explain how it fits into the more general framework introduced so far. 
We conclude by comparing the effectiveness of the different relaxations on numerical experiments on clustering in Section \ref{sec:experimental_results}.

\section{Mass transport and the Wasserstein metric}\label{sec:preliminaries}

In this section, we briefly introduce the problem of optimal mass transport. 
We refer the reader to \cite{Villani2003,Rachev1998-I,Rachev1998-II,Villani2008} for a survey of the subject.
The original formulation of the problem goes back to G. Monge in 1781 \cite{Monge1781}, while the modern formulation is due to L. Kantorovich in 1942 \cite{Kantorovich1942}.
Below we present the Monge-Kantorovich {\em optimal mass transportation (OMT)} problem restricting our discussion to (finite) discrete spaces.

\subsection*{Optimal mass transport}
Let $\Xc$ and $\Yc$ be finite discrete spaces and $p_X$ and $p_Y$ be probability measures on $\Xc$ and $\Yc$, respectively.
Let $c: \Xc \times \Yc \mapsto \mathbb{R} \cup \{ + \infty \}$ be a 
cost function, i.e. $c(x_i, y_j)\geq 0$ represents the transportation cost of transferring one unit of mass from $x_i\in\Xc$ to $y_j\in \Yc$. 
Let $\pi: \Xc \times \Yc \mapsto [0,1]$ be a transference plan (informally, $\pi(x_i,y_j)$ measures the amount of mass transferred from location $x_i$ to location $y_j$).
The OMT problem is
to minimize the total transportation cost over the set of (joint) probability measures $\pi$ with given marginals $p_X$ and $p_Y$ and it is as follows:
\begin{eqnarray*}
\underset{\pi}{\min} & \sum_{i,j}   c(x_i,y_j) \pi(x_i,y_j)   \\
{\text{subject to}} &  \sum_{j} \pi(x_i,y_j) = p_X(x_i), \; \forall i  \\
&  \sum_{i} \pi(x_i,y_j) = p_Y(y_j), \; \forall j \label{probl:Kant_probl_scal_c2}\\
&  \pi(x_i,y_j)\geq 0, \; \forall i,j.
\end{eqnarray*}
The following probabilistic interpretation is standard. If $X, Y$ are random variables taking values on $\Xc$ and $\Yc$ with probability
distributions $p_X$ and $p_Y$, respectively, the OMT problem is to minimize the expectation $\Ebb_\pi \left[c(X,Y)\right]$
over all admissible joint distributions $\pi$ of $(X,Y)$.

We denote by $\Pi$ the matrix associated to the transference plan $\pi$ and by $C$ the matrix associated to the transference cost, i.e.,
$$
\Pi = \left[\pi(x_i,y_j)\right], \quad C = \left[c(x_i,y_j)\right], \quad
$$
for $i=1, \dots, \left| \Xc \right|$, $j=1, \dots, \left| \Yc \right|$, with $\left| \cZ \right|$ denoting the cardinality of the set $\cZ$.
The OMT problem can be expressed in matrix notation as follows:
\begin{eqnarray}
\underset{\Pi\in \Mc(p_X, p_Y)}{\min} & \trace\left(C^T \Pi \right)
\end{eqnarray}
where
\[
\Mc(p_X, p_Y):=\left\{\Pi \mid \Pi  \, \mOne = p_X, \Pi^T  \mOne = p_Y,  \Pi \geq 0   \right\}\,
\]
and $\mOne$ is a vector of ones of suitable dimension.

\subsection*{Wasserstein metric}
Consider the case where $\cZ$ is a metric space with metric $d(\cdot,\cdot)$, and $\cX,\cY\subseteq\cZ$,
and consider the problem of optimal transport between two probability measures as before.
When the transportation cost $c(\cdot,\cdot)$ is equal to  $d(\cdot,\cdot)^q$ with $q> 0$, the OMT induces a metric on the space of probability measures having finite $q$th-moments
\cite[Chapter 7]{Villani2003}. Herein, we are interested in the case where $q=2$.
More specifically, we view $\Xc,\Yc$
as sets of points in a Euclidean space. The Euclidean metric induces a metric on $\cZ:=\Xc\cup\Yc$ and thereby a cost $c(\cdot,\cdot)$ so that
\[
C_{i,j}=\|z_i-z_j\|^2.
\]
The optimal transport cost
\begin{align}\label{def:Wasserstein_distance}
d_{W_2}(p_1,p_2) := \underset{\Pi\in \Mc(p_1, p_2)}\min\trace(C^T \Pi),
\end{align}
where $\Pi$ is the joint probability on $\cZ\times\cZ$ as before,
gives rise to the \emph{$2$-Wasserstein metric} between $p_1$ and $p_2$:
\[\sqrt{d_{W_2}(p_1,p_2)}.\]
For the case where $\cX,\cY\subseteq \cZ$ are not necessarily equal and
for $p_1=p_X$ having support on $\cX$ and $p_2=p_Y$ having support on $\cY$,
$\Pi$ will have support on $\cX\times\cY$.  Therefore the optimization in \eqref{def:Wasserstein_distance} can be carried out with $\Pi$ restricted to be a probability distribution on $\cX\times\cY$ and $C$ restricted to correspond to distances between points in these two spaces, i.e., $C_{i,j}=\|x_i-y_j\|^2$.
			

\section{The cardinality penalty \& relaxations}\label{sec:cardinality_penalized_probl_and_relax}

We now return to considering the optimization problem in \eqref{probl:OMT_penalized_Wass_dist}
specializing $ \mathcal{I}(\cdot)$  to be the cardinality function $\card(\cdot)$ giving the number of nonzero entries of the argument. We
propose alternative convex relaxations that are suitable to the case where the optimization variable is a probability vector.

We first rewrite problem \eqref{probl:OMT_penalized_Wass_dist}, namely,
	\begin{eqnarray}
	\min_{\substack{p_1 \geq 0\\ \mathds{1}^T p_1 = 1 }} d_{W_2}\left(p_0,\,p_1\right) + \lambda {\rm card} \left(p_1\right),
	 \label{probl:OMT_clustering_matr_Wass_dist_penalized}
	\end{eqnarray}
in terms of the transportation plan $\Pi$. Indeed, from \eqref{def:Wasserstein_distance} and since
\[
p_1=\Pi^T \mOne,
\]
problem \eqref{probl:OMT_clustering_matr_Wass_dist_penalized} can be rewritten as
\begin{eqnarray}
\min_{\substack{\Pi \geq 0\\
\Pi  \, \mathds{1} = p_0}} & \trace \left(C^T \Pi \right)+ \lambda {\rm card}\left(\Pi^\top \mathds{1}\right).  \label{probl:OMT_clustering_matr_no_p1}
\end{eqnarray}
The optimal value will be denoted by $J^{\rm opt}$.

Cardinality penalized problems are in general NP--hard to solve, being combinatorial in nature. 
The cardinality function is nonconvex and is usually replaced by the $\ell_1$-norm which is a convex surrogate \cite[Chapter 6]{BoydVand2004}. Evidently, such a relaxation is not applicable here since we are dealing with probability vectors. Below, in \ref{sec:GLasso_relaxation} and  \ref{sec:y_relaxation}, we propose two alternative relaxations that are applicable to our setting and, in \ref{sec:l_inf_relaxation}, we discuss an additional relaxation which has recently been proposed in \cite{ElGhaoui2012NIPS}.

\subsection{Rank regularization and sum--of--norms relaxation}\label{sec:GLasso_relaxation}

The main idea underlying the relaxation proposed below
is to express the cardinality penalty in terms of the rank of a certain linear map of the transportation plan $\Pi$.
To this end, we denote by
$e_i\in{\mathbb R}^N$ the standard unit $N$-vector with $1$ in the $i$th entry and with $E_i$ the single--entry diagonal matrix with a $1$ in position $(i,i)$. 
Moreover, we denote by $\Pi_i$ the $i$th column of $\Pi$. 
We introduce the map $F$ that associates to each transportation plan $\Pi$ the rectangular block--diagonal matrix with diagonal block-entries the columns of $\Pi$,
\begin{equation}\label{eqn:Pit}
F(\Pi) := \sum_{i=1}^N E_i \otimes (\Pi \, e_i)  = 
\bmat
\Pi_1 & 0 & \dots & 0\\
0& \Pi_2 &  \dots & 0\\
\vdots & \vdots &  \\
0& 0 & \dots & \Pi_N 
\emat\,,
\end{equation} 
where $\otimes$ stands for the Kronecker product. Whenever it is clear from the context we simplify the notation and denote $F(\Pi)$ simply by $F$.
The matrix $F^T F$ is diagonal with $i$th diagonal entry the scalar product 
$\left\langle \Pi_i, \Pi_i\right\rangle$. Since $\left\langle \Pi_i, \Pi_i\right\rangle=0$ if and only if $\Pi_i=0$, we have  
\begin{eqnarray*}
\card(\Pi^T \mOne)&=&\rank\left(F^T F\right)\\
&=&\rank\left(F \right)
\end{eqnarray*}
and \eqref{probl:OMT_clustering_matr_no_p1} now becomes 
\begin{align}\label{prob:rank}
\min_{\substack{\Pi \geq 0\\
\Pi  \, \mathds{1} = p_0}} &  \trace\left(C^T \Pi \right)+ \lambda\, \rank\left(F(\Pi)\right)  \,.
\end{align}

Denote by $||F||_2$ the spectral norm of $F$ and by $||F||_*$ its nuclear norm. 
Due to the particular structure of $F$, $F^TF$ is diagonal and the cardinality of $\diag(F^TF)$ coincides with the rank of $F$.
Then, $||F||_*$, which is defined as the sum of the singular values, is simply 
\begin{eqnarray*}
||F||_*&=&\|\left(\diag(F^TF)\right)^\frac12\|_1\\
&=&\sum_{i=1}^N ||\Pi_i||_2.
\end{eqnarray*}
Utilizing the well-known fact that the nuclear norm represents the  convex envelope of the rank function on a (bounded set) of matrices
(Fazel {\em et al.} \cite{Fazel2001}), we obtain the following result.  

\begin{proposition}\label{thm:tightest_lower_bound}
{\em The convex envelope of 
$\rank(F(\Pi))$ on the set $\left\{\Pi \, | \, \Pi  \, \mOne = p_0,\, \Pi \geq 0 \right\}$
is
\[
\phi(\Pi) = \frac{1}{||p_0||_2} \sum_{i=1}^N ||\Pi_i||_2\,.
\]}
\end{proposition}
\vspace{3mm}
\begin{proof} 
Since by \cite[Theorem 1]{Fazel2001}, it holds that $||M||_*/a$ is the \emph{convex envelope} of $\rank(M)$ over the set $\left\{M \, \mid \,||\,M||_2 \leq a\right\}$, 
the only thing that remains to prove is that 
$||F||_2$  
is bounded by $\|p_0\|_2$ on the feasible set. 
Indeed, for every $\Pi \in \left\{\Pi \, | \, \Pi  \, \mOne = p_0,\, \Pi \geq 0 \right\}$, $F(\Pi)$ given by \eqref{eqn:Pit},  it holds that
\begin{align*}
||F||_2 & = \underset{i}{{\rm max}} \sqrt{{\rm eig}\left(F^T F\right)} \\
&= \underset{i}{{\rm max}} \, {\rm eig}\left\{\bmat ||\Pi_1||_2 & 0& \dots & 0\\
0 & ||\Pi_2||_2& \dots & 0\\
\vdots & \vdots &  & \\
0 & 0& \dots & ||\Pi_N||_2\\
\emat\right\}\\
&=\underset{i}{{\rm max}} \left\{||\Pi_i||_2\right\}
\leq \|\sum_i \Pi_i\|_2=\|p_0\|_2\,,
\end{align*}
as claimed. 
\end{proof}

\newcommand{\gl}{{\rm glasso}}
It follows that the rank--penalized problem \eqref{prob:rank} can be relaxed into 
\begin{eqnarray}\label{probl:OMT_clustering_matr_no_p1_relax}
\min_{\substack{\Pi \geq 0\\
\Pi  \, \mathds{1} = p_0}} & \trace \left(C^T \Pi \right)+ \frac{\lambda}{\|p_0\|_2} \sum_{i=1}^N \left\|\Pi_i\right\|_2   \,. 
\end{eqnarray}
where the sum--of--norms penalty 
is a Group--Lasso--type penalty \cite{YuanLin2005}  
with the groups given by the columns of the transference plan $\Pi$.

\subsection{Integer programming and fractional relaxation}\label{sec:y_relaxation}

In this section, an alternative relaxation of Problem \eqref{probl:OMT_clustering_matr_no_p1} is presented. 
To this aim, we introduce the indicator function 
\begin{equation*}
 y_i=\left\{
       \begin{array}{ll}
         1, & \hbox{if $\Pi_i\neq 0$} \\
         0, & \hbox{otherwise}
       \end{array}
     \right.
\end{equation*}
whose entries reflect the sparsity pattern of the columns of $\Pi$. 
Then, clearly,
\[
\card(\Pi^T \mOne)= \|y\|_1,
\]
Moreover, since the the columns of $\Pi$ must sum up to $p_0$, the following inequality holds  
\[
\Pi \leq p_0 y^T.
\]
This leads to the following equivalent formulation of \eqref{probl:OMT_clustering_matr_no_p1}
\begin{align}\label{prob:y}
\underset{\Pi, y}{{\rm minimize}} ~~~~& \trace\left(C^T \Pi \right)+ \lambda\|y\|_1 \\
\text{subject to } ~~~~&\Pi  \, \mOne = p_0, \nonumber\\
&\Pi \geq 0,\nonumber\\
&\Pi \leq p_0 y^T,\nonumber\\
&y_i \in \{0, 1\}. \nonumber 
\end{align}
Problem \eqref{prob:y} is not convex due to the integer constraints $y_i\in \{0, 1\}$. 
A standard relaxation of an integer program is 
\begin{align}\label{prob:y_relax}
\underset{\Pi, y}{{\rm minimize}} ~~~~& \trace\left(C^T \Pi \right)+ \lambda\|y\|_1 \\
\text{subject to } ~~~~&\Pi  \, \mOne = p_0, \nonumber\\
&\Pi \geq 0,\nonumber\\
&\Pi \leq p_0 y^T,\nonumber\\
&y_i \in [0, 1],  \nonumber
\end{align}
where $y_i\in\{0,1\}$ is relaxed into $y_i\in[0, 1]$. 
This formulation is similar to the so called {\em facility location} problem in Operations Research \cite{Daskin1995,DreznerHamacher2001},
where typically the matrix $\Pi$ is Boolean ($(0,1)$-entries) and $p_0$ has integer entries, and a number of works has been devoted to the design of specialized LP--rounding algorithm for reconstructing integer solutions starting from a solution of a relaxed problem (see e.g. \cite{LinVitter1992,CharikarTardosShmoys1999,CharikarTardosShmoys2002,JainVazirani2001}).

In section \ref{sec:experimental_results}, the relaxation \eqref{prob:y_relax} will be compared with the sum--of--norms relaxation \eqref{probl:OMT_clustering_matr_no_p1_relax} 
on a clustering application.

\subsection{Relaxation inverse of the $\ell_\infty$-norm }\label{sec:l_inf_relaxation}

The relaxation discussed below is a special case of a problem that is addressed by Pilanci {\em et al.} in \cite{ElGhaoui2012NIPS}
where they consider sparse minimizers of general convex functions on probability simplices. 
The key idea in \cite{ElGhaoui2012NIPS} is to utilize
the inverse of the $\ell_\infty$-norm of the probability vector (here, $p_1$) as a surrogate for the cardinality.
Indeed, in general,
\[
\left\|p_1\right\|_1 \leq 
{\rm card}\left(p_1\right) \left\|p_1\right\|_\infty,
\]
and since $p_1$ is a probability vector,
\[
\frac{1}{\left\|p_1\right\|_\infty} \leq {\rm card}\left(p_1\right)\,.
\]
Problem \eqref{probl:OMT_clustering_matr_no_p1} can thus be relaxed to
\begin{eqnarray}\label{probl:OMT_clustering_matr_ElGhaoui}
\min_{\substack{\Pi \geq 0\\
\Pi  \, \mathds{1} = p_0}} & \trace\left(C^T \Pi \right)
+ \, \frac{\lambda\;}{\left\|\Pi^T \mOne\right\|_\infty}\,\, .
\end{eqnarray}

Note that \eqref{probl:OMT_clustering_matr_ElGhaoui} is still not a convex problem. 
Nevertheless, it can be solved exactly by 
using the following $N$ convex programs (see \cite[Proposition 2.1]{ElGhaoui2012NIPS})
\begin{eqnarray}\label{probl:OMT_clustering_matr_ElGhaoui_convex}
\min_{i=1, \dots,N} \left\{ \min_{\substack{\Pi \geq 0\\\Pi  \, \mathds{1} = p_0}} \trace\left(C^T \Pi \right) + \frac{\lambda}{[\Pi^T \mOne]_i} \right\}  
\end{eqnarray}
where $\left[\Pi^T \mOne \right]_i$ denotes the $i$th component of the vector $\Pi^T \mOne$.

In \cite{ElGhaoui2012NIPS}, this relaxation has been applied 
to estimate the (sparse) coefficients vector of a Gaussian mixture in the exemplar based convex clustering framework of \cite{Lashkari2007}. 
 
The relaxation based on the $\ell_\infty$--norm will be compared with the sum--of--norm relaxation \eqref{probl:OMT_clustering_matr_no_p1_relax} 
and the relaxation based on the integer programming formulation \eqref{prob:y_relax} in Section \ref{sec:experimental_results}.

\section{Clustering via OMT}\label{sec:clustering_via_OMT}

In this section, we describe how ideas from optimal mass transport can be applied for clustering points in a metric space.    
Consider a data set $\Xc = \left\{\xb_1, \dots, \xb_N\right\}\subset \mathbb{R}^d$. 
A common way to address the problem is to select a subset of cluster representatives 
and associate points to these representatives in such a way to minimize a given cost functional, 
which is usually the sum of the square distances between points and associated cluster representatives. 
Here we observe that the problem of optimally selecting a subset of data points as cluster centers, can be seen as an optimal mass transportation problem where
a sample distribution $p_0$ is associated to the data points in $\Xc$ and a distribution $p_1$ is associated to the cluster centers.  
Cluster centers are chosen in such a way to minimize the optimal transportation cost between $p_0$ and $p_1$. 
In this, $p_1$ is to be determined based on the requirements that its support consists of a few points (few cluster representatives). 
This leads to a \emph{convex} clustering scheme where the optimal transportation plan $\Pi^{\rm opt}$ can be computed by solving %
one of the proposed relaxations of \eqref{probl:OMT_clustering_matr_no_p1}  and 
clustering is achieved according to the following rules: 
\begin{itemize}
		\item Choice of cluster representatives: the point $\xb_j$ is a representative of a cluster if  $\exists$ a point $\xb_i$ such that $\Pi^{\rm opt}_{ij}>\Pi^{\rm opt}_{i\ell}$ for all $\ell \neq j$,
		\item Association of points to cluster representatives: assign the point $\xb_i$ to the cluster representative $\xb_j$ if $\Pi^{\rm opt}_{ij}>\Pi^{\rm opt}_{i\ell}$ for all $\ell \neq j$.
\end{itemize}

\section{Experimental results}\label{sec:experimental_results}

Following the scheme of the previous section, we now compare the relaxations introduced so far on a clustering example. 
In particular, we consider synthetic data in $\mathbb{R}^2$ sampled from $4$ Gaussian distributions with means 
\begin{align*}
\left\{
\left(0,5\right), \left(-5\frac{\sqrt{3}}{2},-\frac{5}{2}\right), \left(5\frac{\sqrt{3}}{2},-\frac{5}{2}\right),\left(8,2\right)\right\}
\end{align*}
respectively, and common covariance given by 
$$
\bmat 0.8 & 0 \\ 0 & 0.8\emat\,.
$$

\begin{figure*}[htp]
\centering
\subfloat[][$\lambda=750$]{\includegraphics[width=0.3\textwidth]{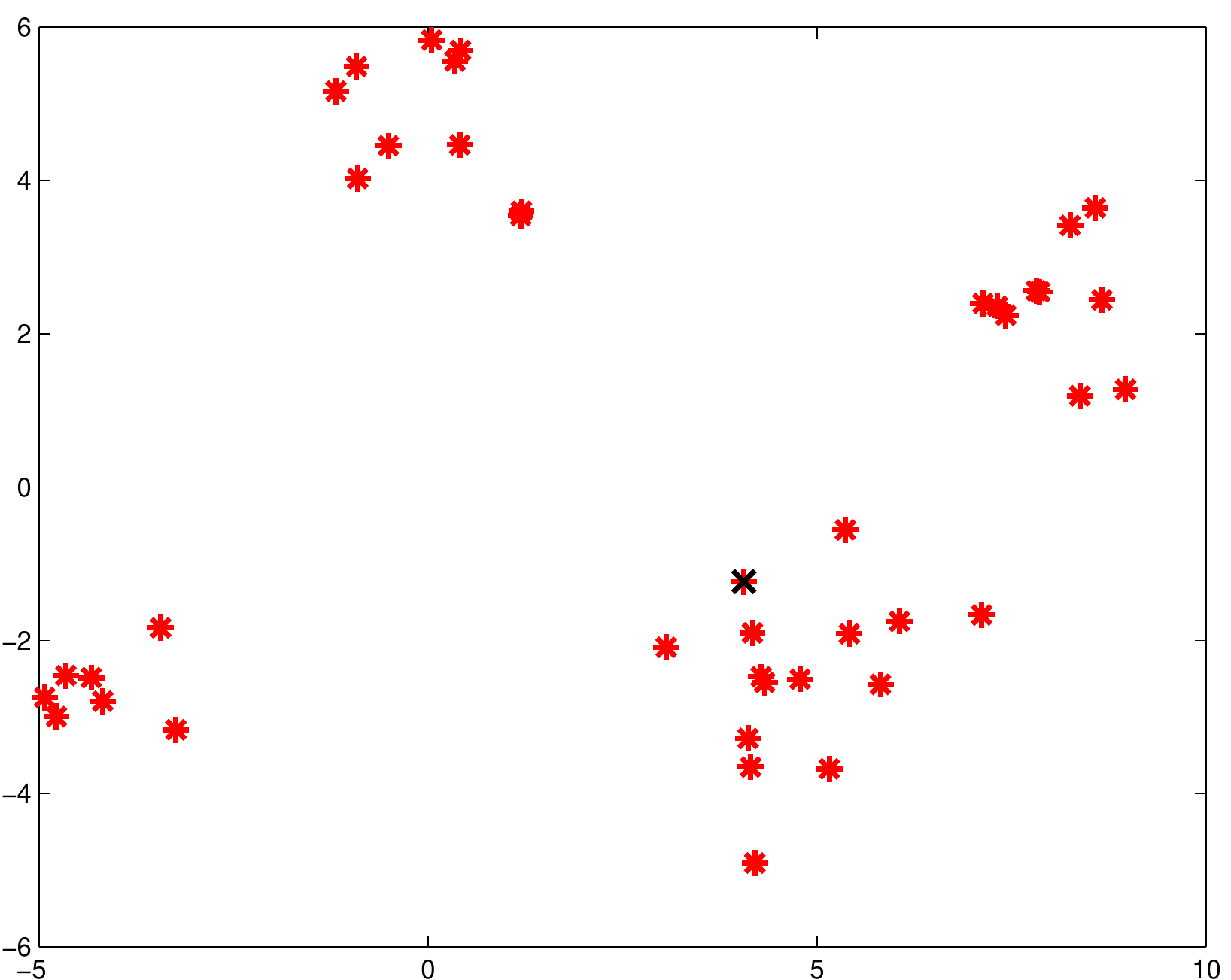}\label{fig:FourCloud_M_GL_lambda750_var8_s7}}\hspace{2mm}
\subfloat[][$\lambda=220$]{\includegraphics[width=0.3\textwidth]{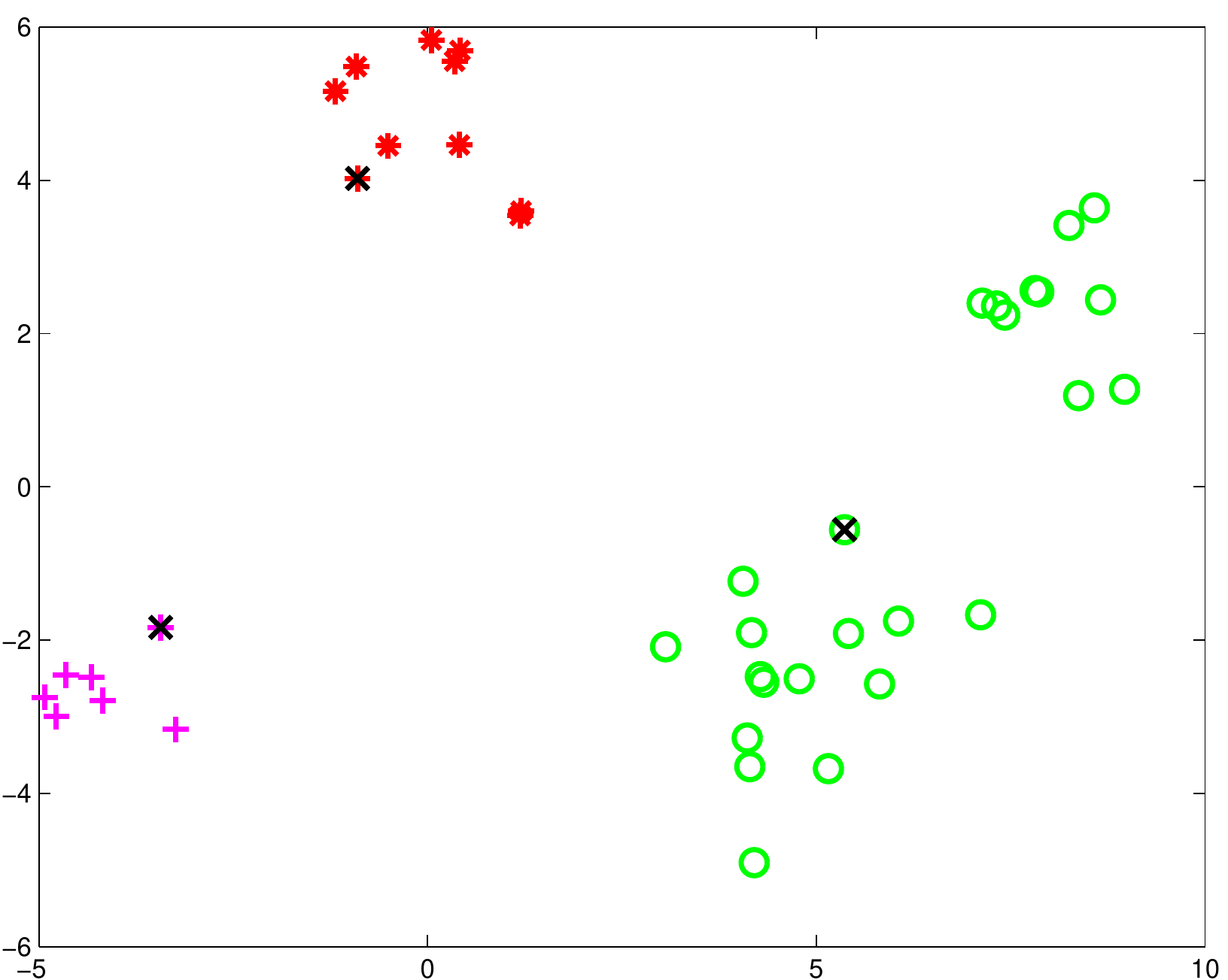}\label{fig:FourCloud_M_GL_lambda10_var8_s7}}
\\\subfloat[][$\lambda=100$]{\includegraphics[width=0.3\textwidth]{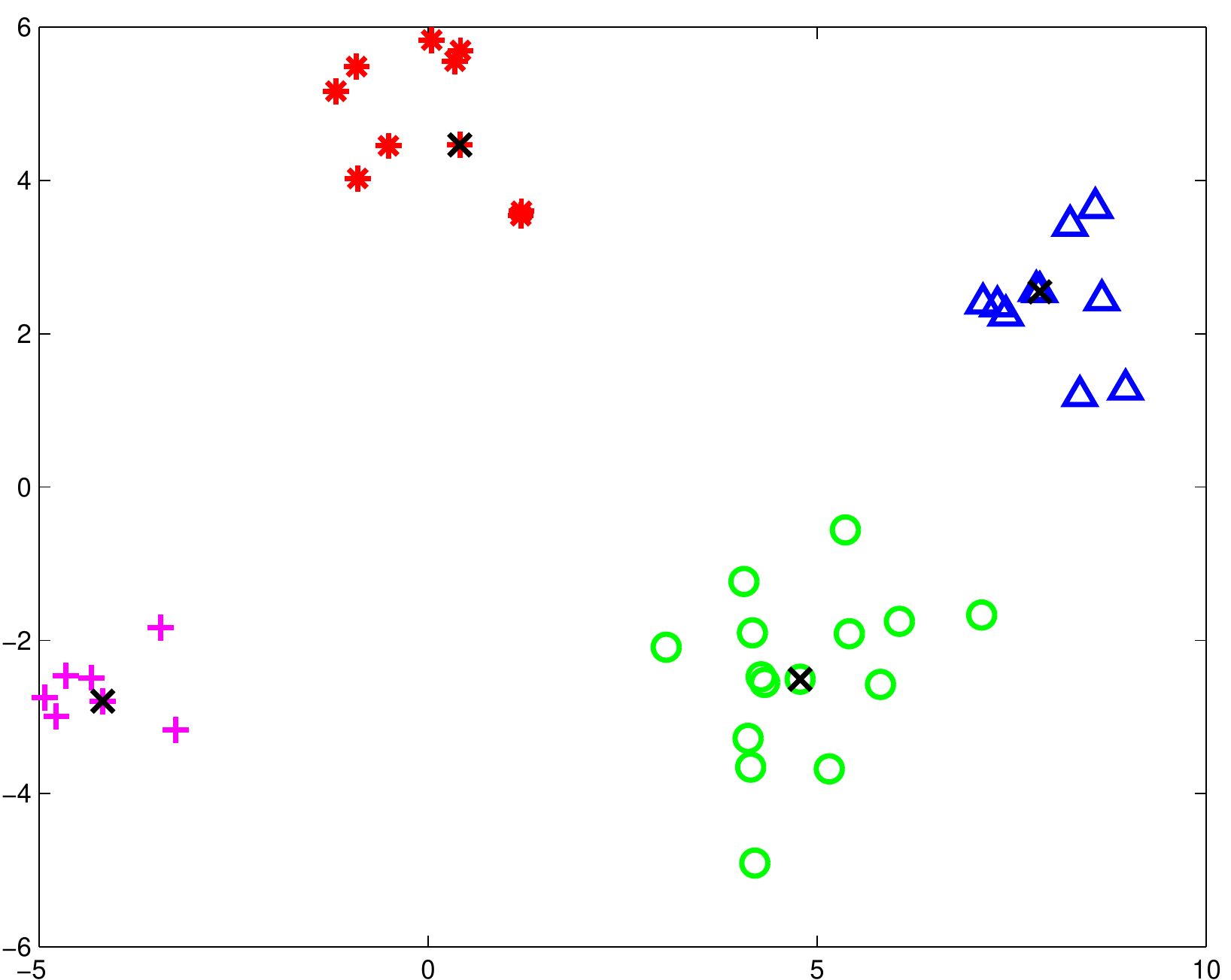}\label{fig:FourCloud_M_GL_lambda100_var8_s7}}\hspace{2mm}
\subfloat[][$\lambda=20$]{\includegraphics[width=0.3\textwidth]{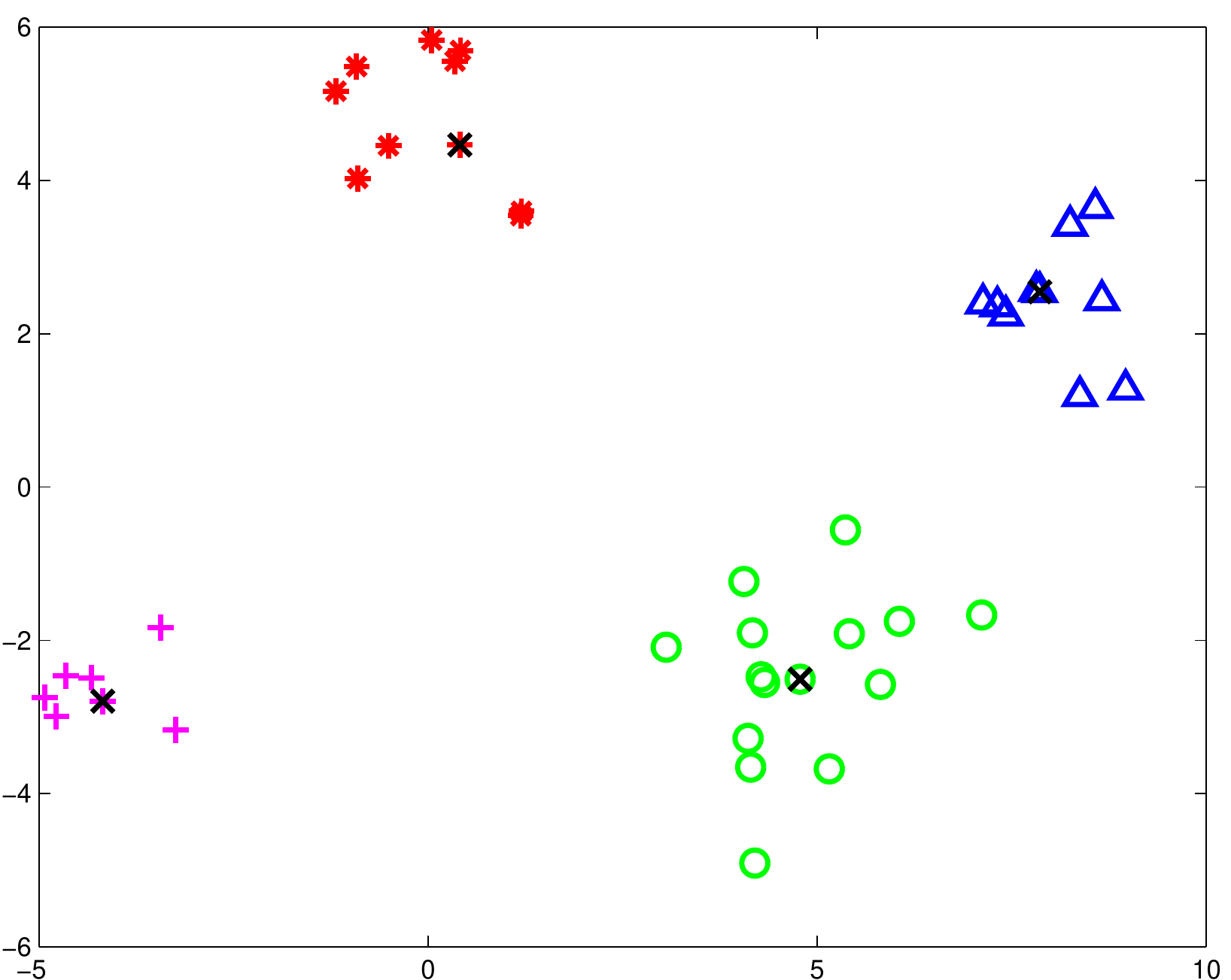}\label{fig:FourCloud_M_GL_lambda20_var8_s7}}
\caption{Output clustering via the solution of the sum--of--norms relaxation \eqref{probl:OMT_clustering_matr_no_p1_relax} for different values of the parameter 
$\lambda$.  
The clusters representatives are denoted by a black cross. Different shapes and colors have been used to denote points belonging to different clusters. }
\label{fig:GL_different_lambda_4clusters}
\end{figure*}

\begin{figure*}[htp]
\centering
\subfloat[][$\lambda=750$]{\includegraphics[width=0.30\textwidth]{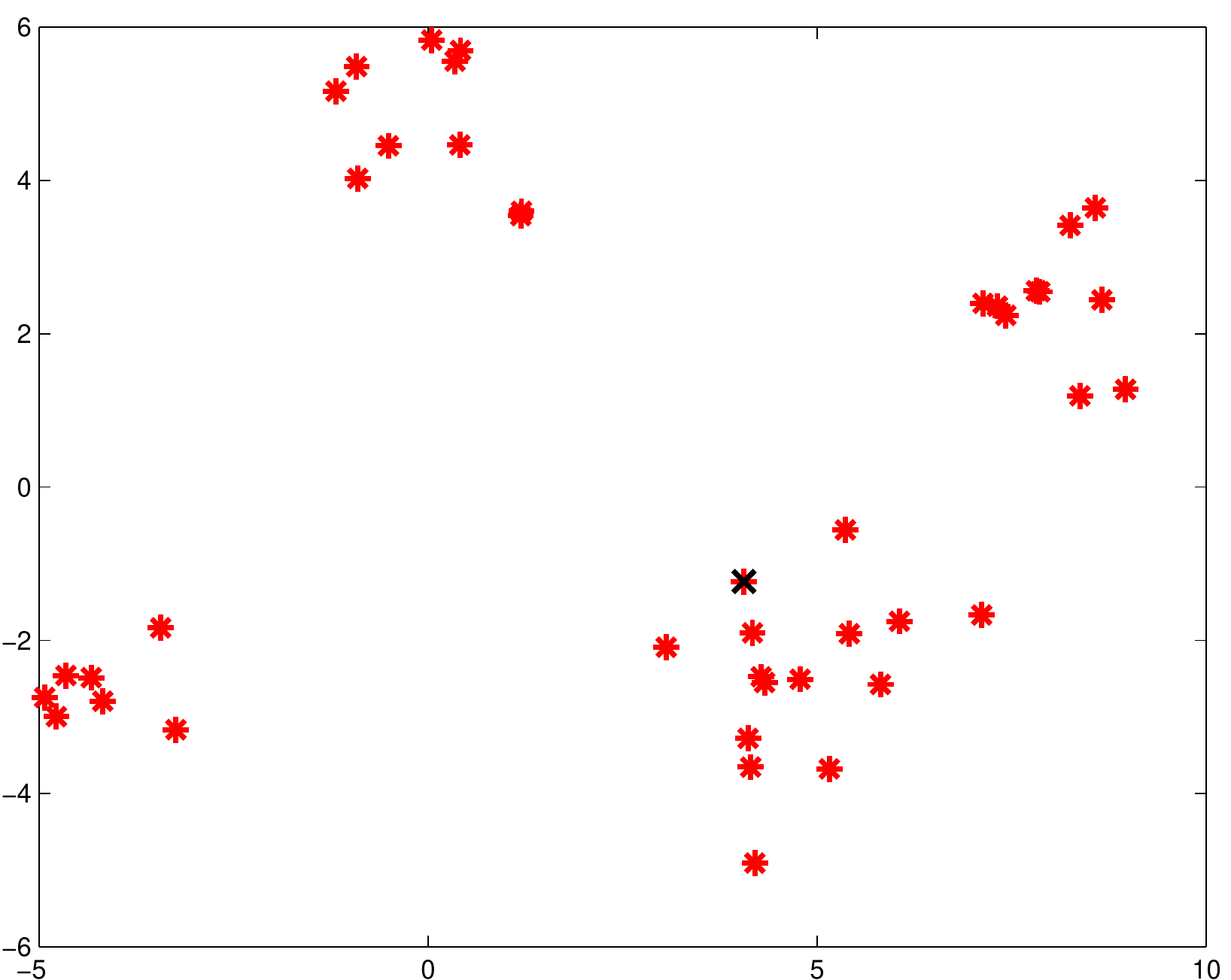}\label{fig:FourCloud_M_LP_lambda750_var8_s7}}\hspace{2mm}
\subfloat[][$\lambda=220$]{\includegraphics[width=0.30\textwidth]{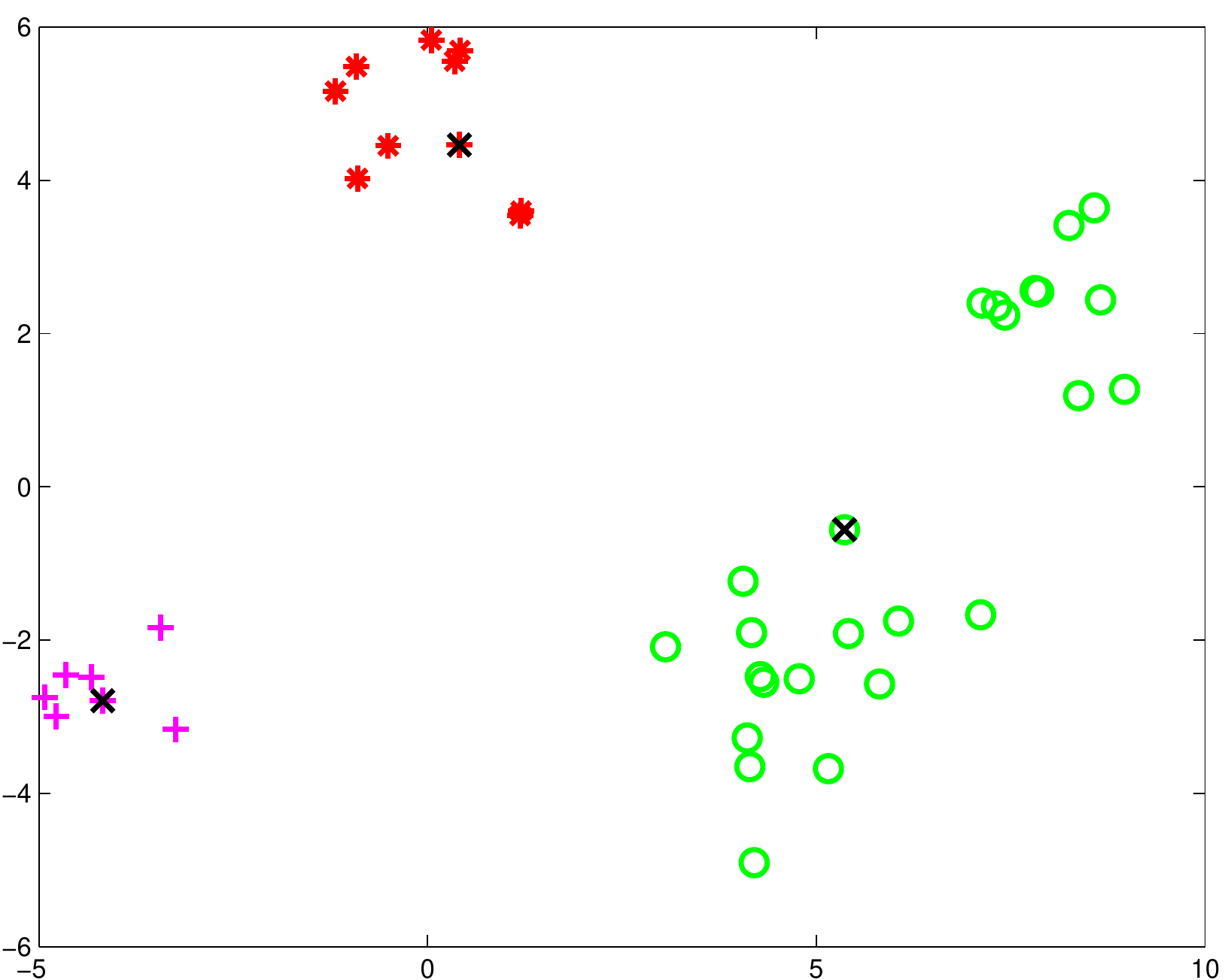}\label{fig:FourCloud_M_LP_lambda10_var8_s7}}
\\\subfloat[][$\lambda=100$]{\includegraphics[width=0.30\textwidth]{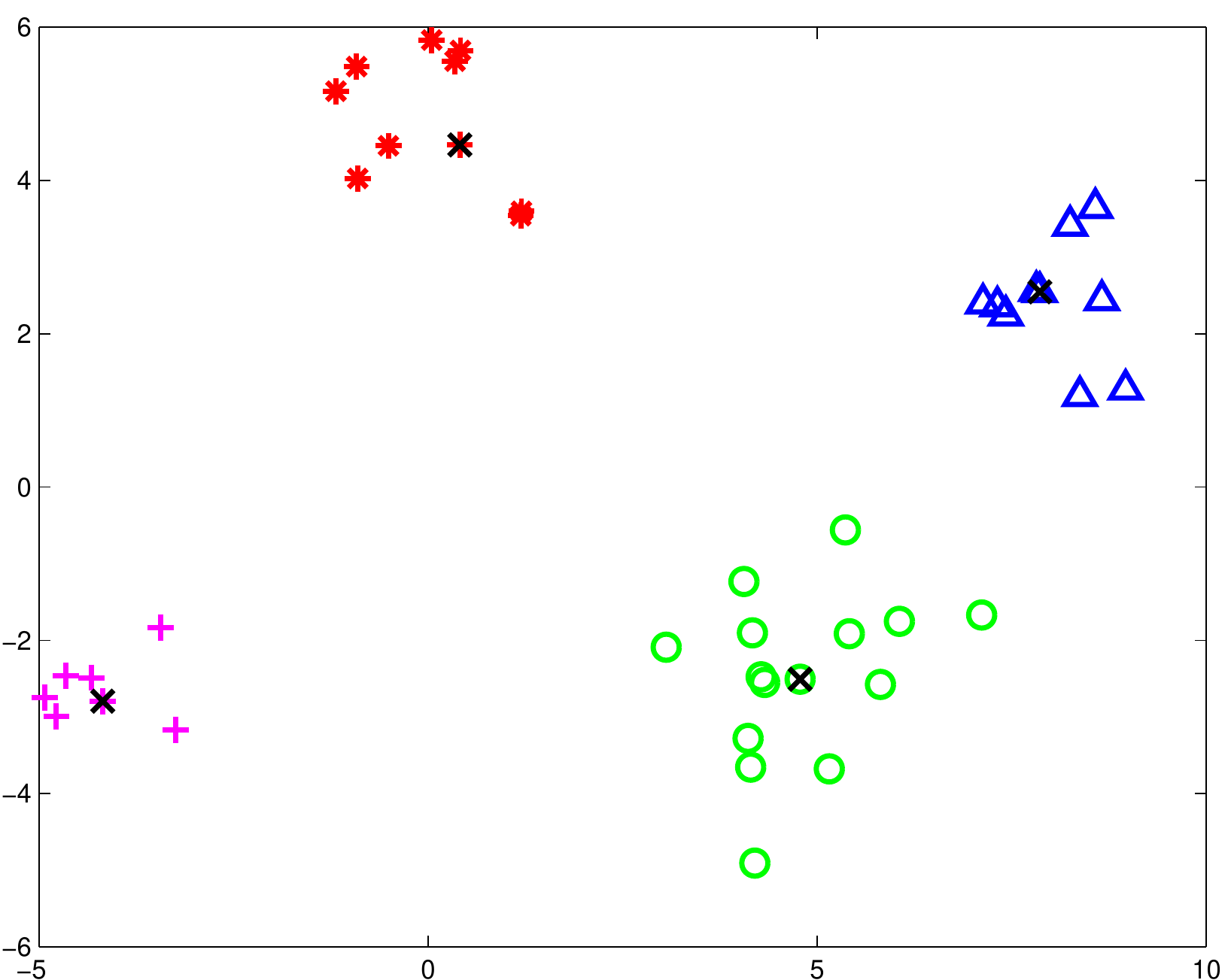}\label{fig:FourCloud_M_LP_lambda100_var8_s7}}\hspace{2mm}
\subfloat[][$\lambda=20$]{\includegraphics[width=0.30\textwidth]{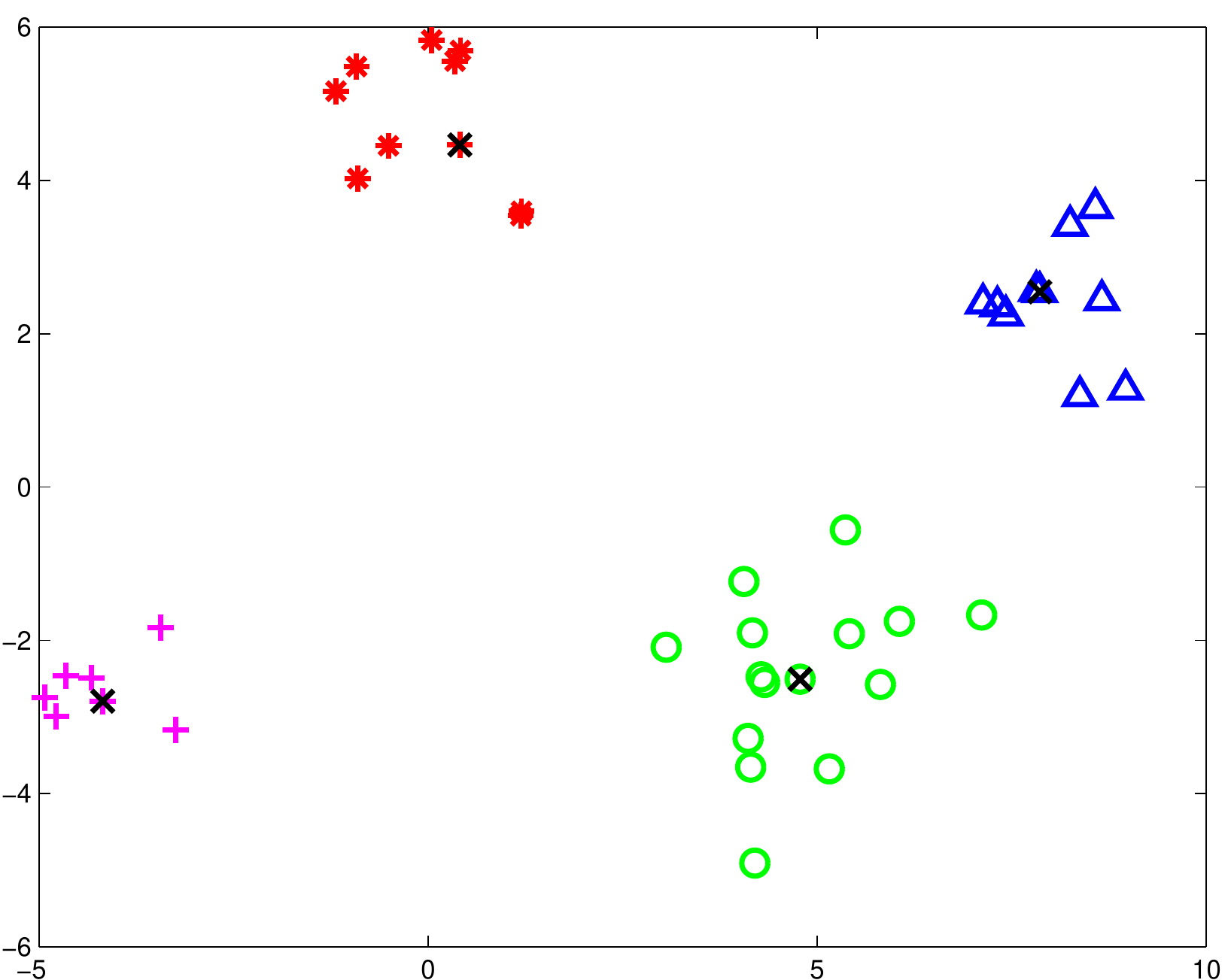}\label{fig:FourCloud_M_LP_lambda20_var8_s7}}
\caption{Output clustering via the solution of the LP formulation \eqref{prob:y_relax} for different values of the parameter 
$\lambda$.  
The clusters representatives are denoted by a black cross. Different shapes and colors have been used to denote points belonging to different clusters. }
\label{fig:LP_different_lambda_4clusters}
\end{figure*}

\begin{figure*}[htp]
\centering
\subfloat[][$\lambda=100$]{\includegraphics[width=0.30\textwidth]{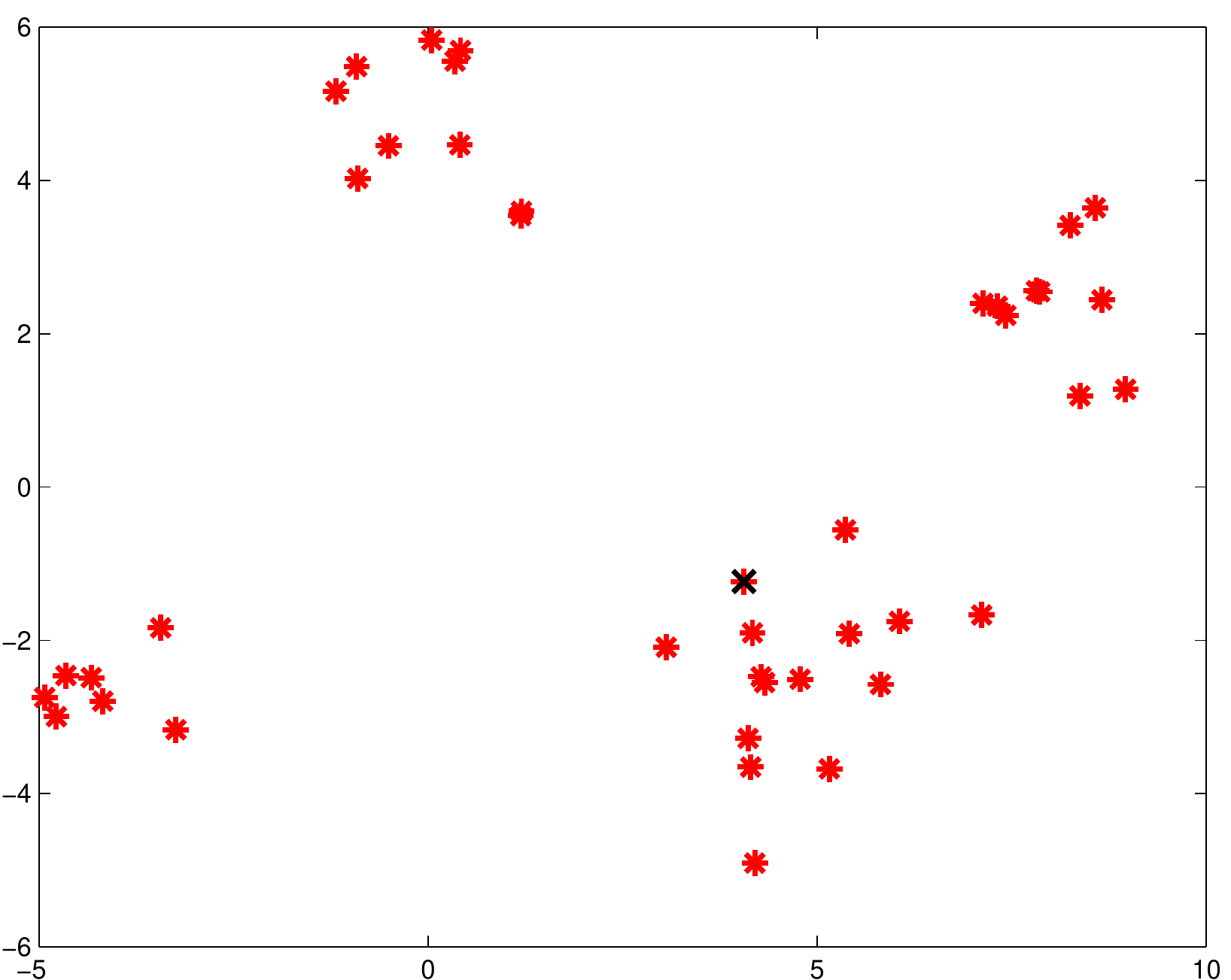}\label{fig:FourCloud_M_inf_lambda100_s7}}\hspace{2mm}
\subfloat[][$\lambda=40$]{\includegraphics[width=0.30\textwidth]{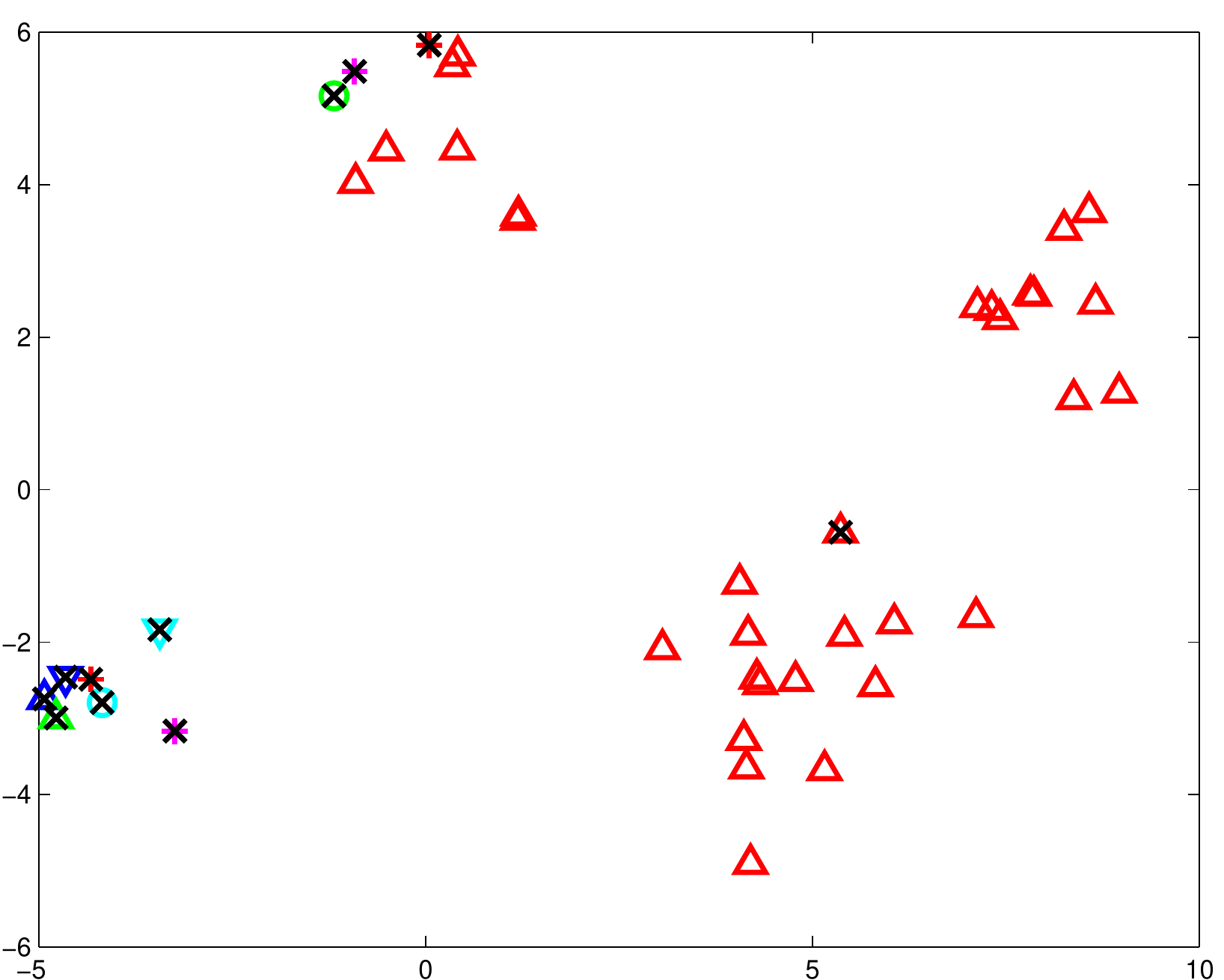}\label{fig:FourCloud_M_inf_lambda40_s7}}
\\\subfloat[][$\lambda=10$]{\includegraphics[width=0.30\textwidth]{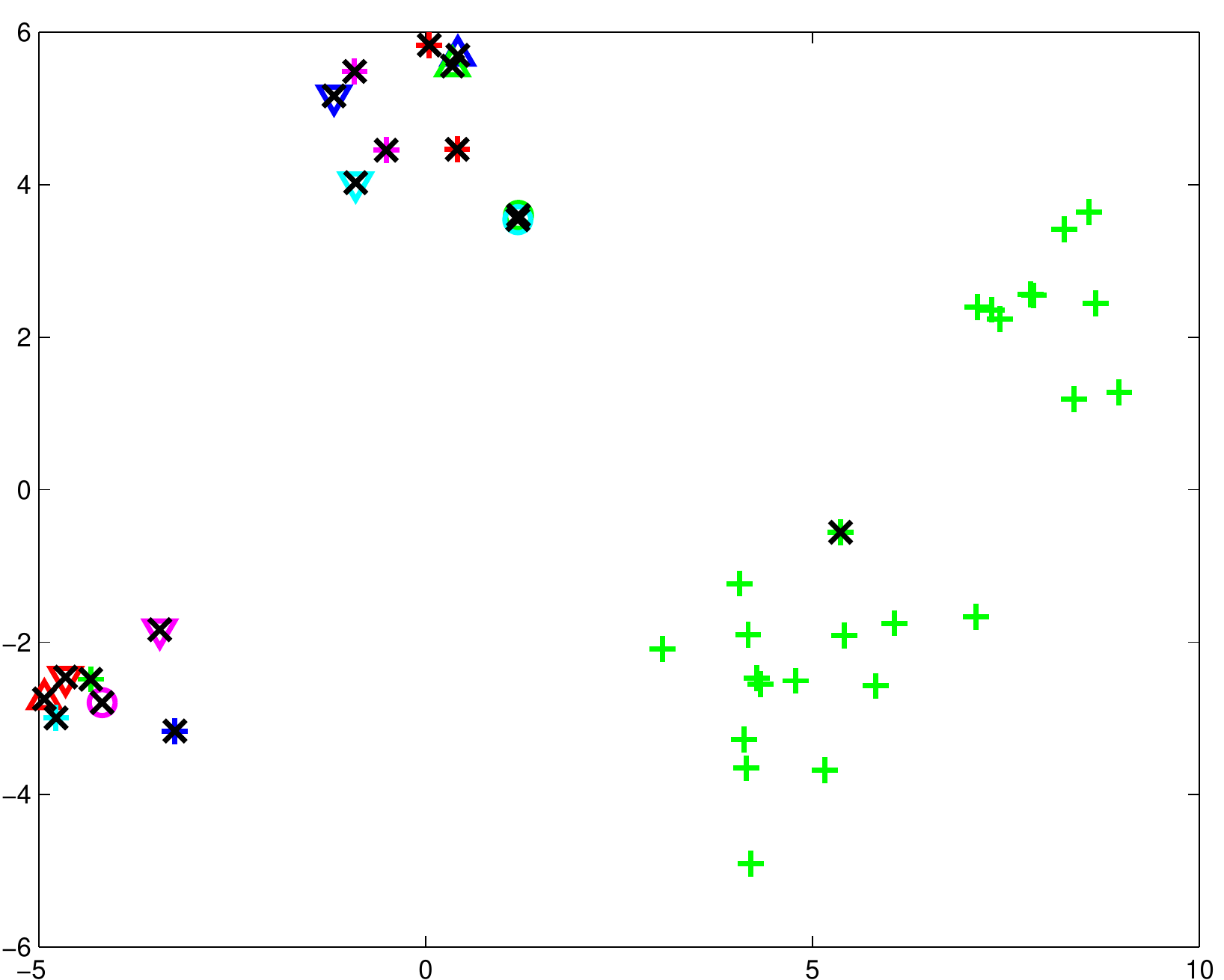}\label{fig:FourCloud_M_inf_lambda10_s7}}\hspace{2mm}
\subfloat[][$\lambda=4$]{\includegraphics[width=0.30\textwidth]{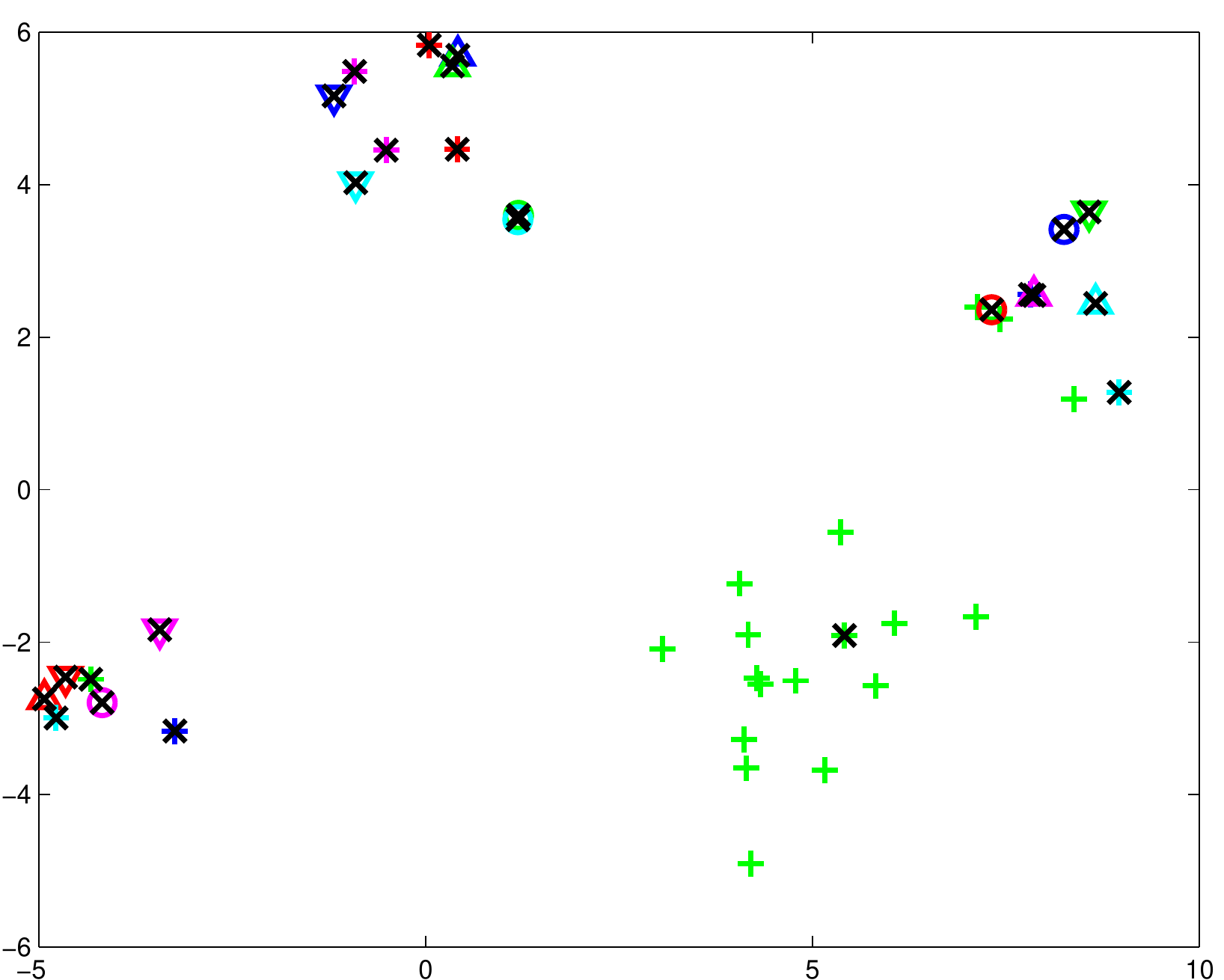}\label{fig:FourCloud_M_inf_lambda4_s7}}
\caption{Output clustering via the solution of the $\ell_\infty$--norm relaxation \eqref{probl:OMT_clustering_matr_ElGhaoui_convex}  for different values of the parameter $\lambda$.  
The clusters representatives are denoted by a black cross. Different shapes and colors have been used to denote points belonging to different clusters. }
\label{fig:inf_different_lambda_4clusters}
\end{figure*}

The clustering obtained by solving \eqref{probl:OMT_clustering_matr_no_p1_relax}, \eqref{prob:y_relax} and \eqref{probl:OMT_clustering_matr_ElGhaoui_convex} for different values of $\lambda$ are shown in Figures \ref{fig:GL_different_lambda_4clusters}, 
\ref{fig:LP_different_lambda_4clusters} and \ref{fig:inf_different_lambda_4clusters}, respectively. 
The clusters representatives are denoted by a black cross while data points belonging to different clusters are denoted by different shapes and colors. 

From the experimental results, we see that both the relaxation based on the sum--of--norms and 
the one based on the integer programming formulation are able to achieve the correct clustering. 
In particular, as it can be seen from panels (c) and (d) in Figures \ref{fig:GL_different_lambda_4clusters} and \ref{fig:LP_different_lambda_4clusters}, 
for values of $\lambda$ approximately between $20$ and $100$, both methods correctly partition the data set into $4$ clusters. 
For $\lambda = 220$ both methods partition the data set into $3$ clusters. 
By further increasing the value of the parameter $\lambda$, data are finally ``grouped'' in a unique cluster, which happens approximately for $\lambda = 750$.  

The clustering produced by optimizing \eqref{probl:OMT_clustering_matr_ElGhaoui_convex} is shown in Figure \ref{fig:inf_different_lambda_4clusters}. 
When applied to our particular problem, 
the penalty based on the $\ell_\infty$--norm is less effective than 
the proposed relaxations in promoting sparsity of the clusters representatives distribution. 
Clustering results for different values of the parameter $\lambda$ are shown in Figure \ref{fig:inf_different_lambda_4clusters}. 
For $\lambda=100$, the data points are ``grouped'' in a unique cluster. 
By decreasing the value of $\lambda$ the method is able to isolate the biggest cluster, 
while all the other points remain unclustered (each point is chosen as the representative of itself). 
There are no values of $\lambda$ for which the penalty based on the $\ell_\infty$--norm is able 
to correctly partition the data set in four clusters.

\begin{figure*}[htp]
\centering
\subfloat[][$\lambda=1500$]{\includegraphics[width=0.30\textwidth]{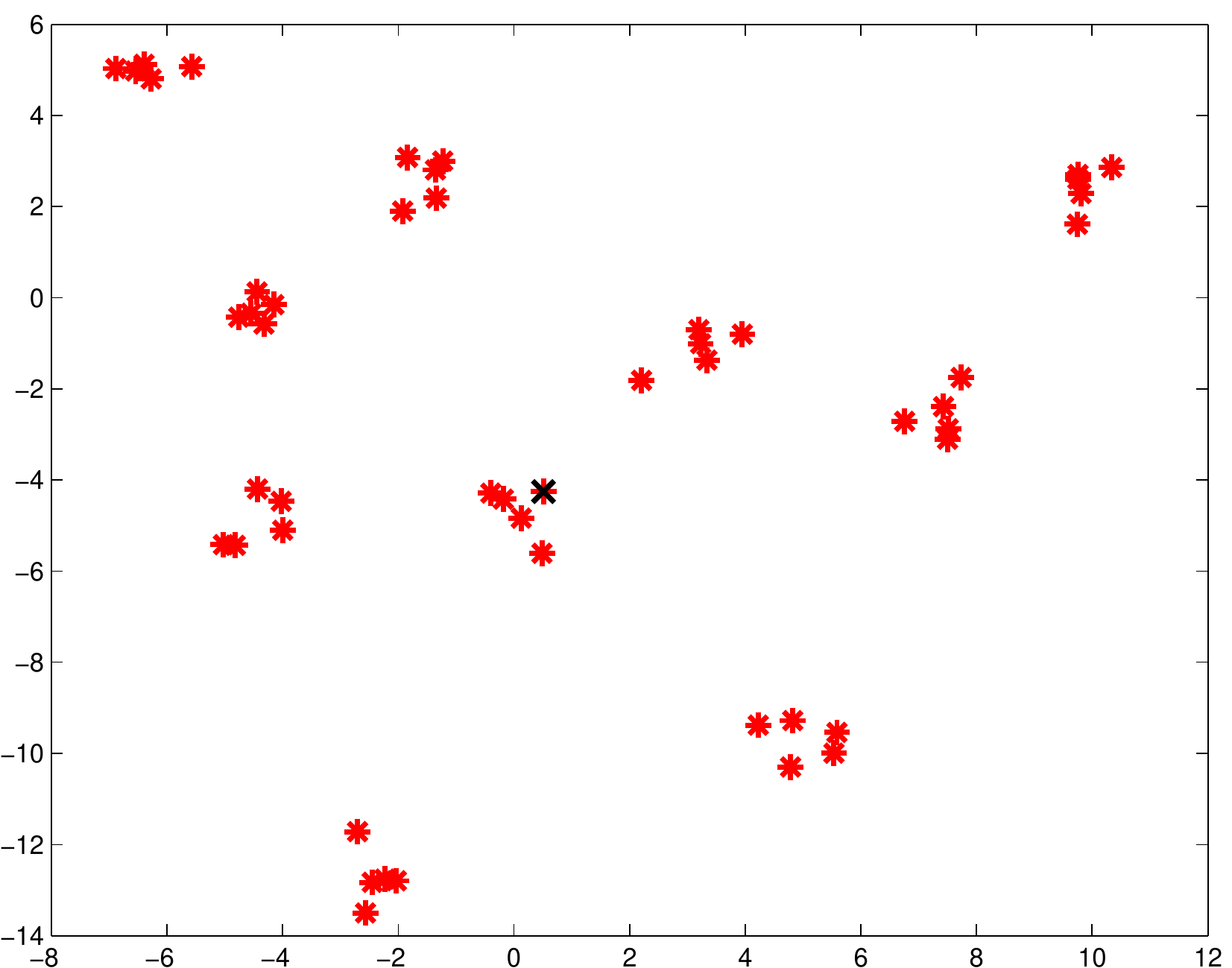}\label{fig:TenClouds_M_GL_lambda500_var2_s2}}\hspace{2mm}
\subfloat[][$\lambda=300$]{\includegraphics[width=0.30\textwidth]{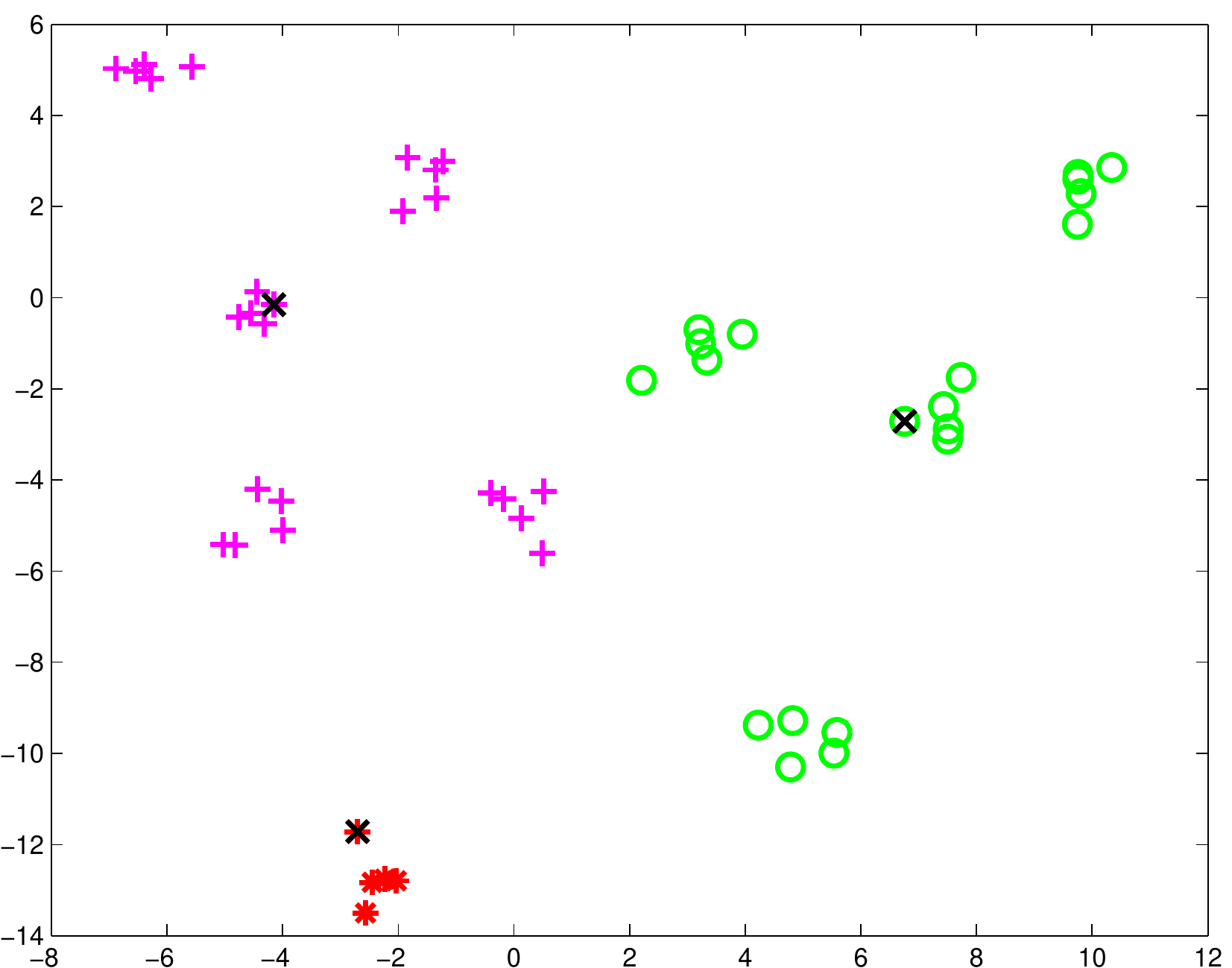}\label{fig:TenClouds_M_GL_lambda100_var2_s2}}
\\\subfloat[][$\lambda=30$]{\includegraphics[width=0.30\textwidth]{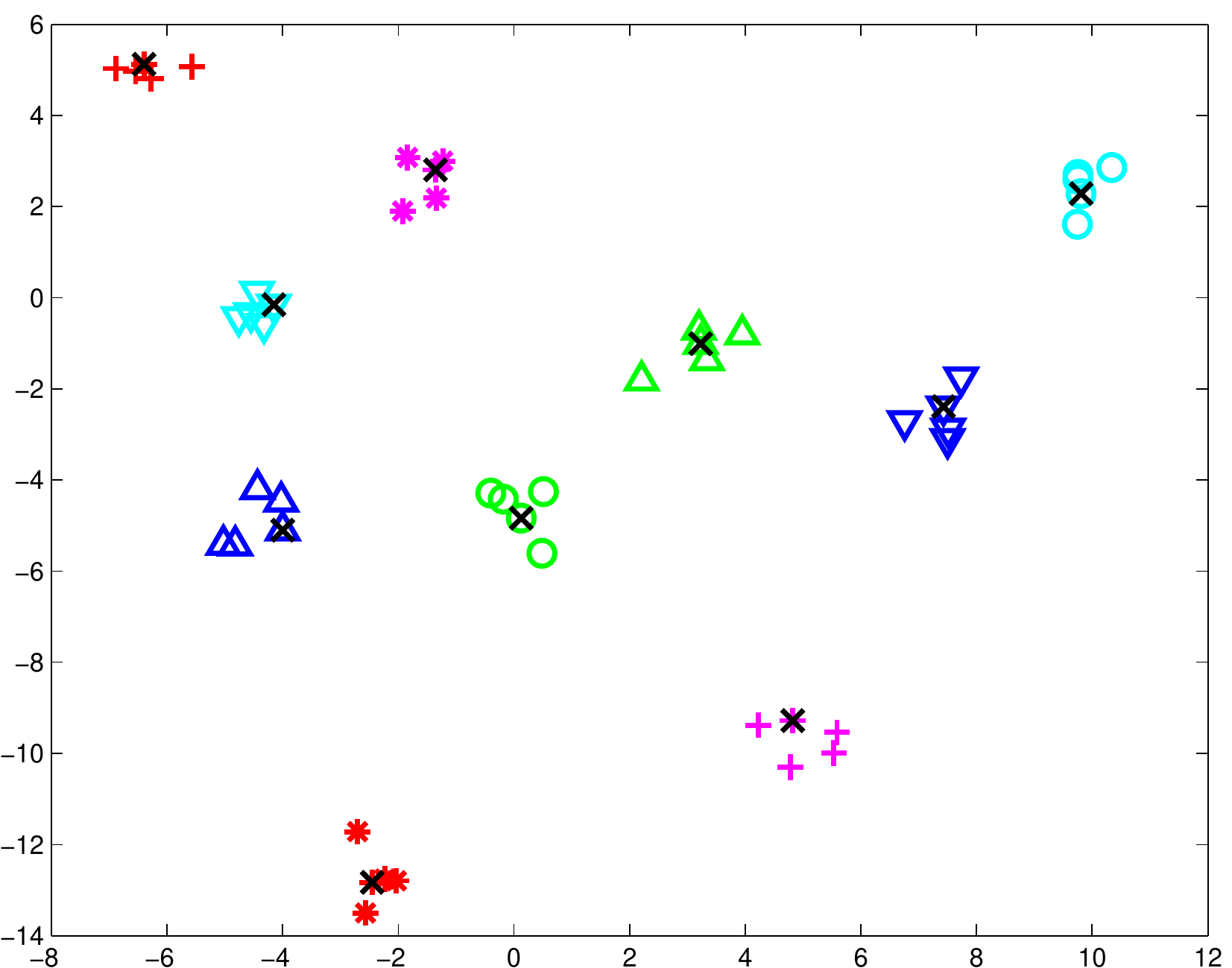}\label{fig:TenClouds_M_GL_lambda50_var2_s2}}\hspace{2mm}
\subfloat[][$\lambda=10$]{\includegraphics[width=0.30\textwidth]{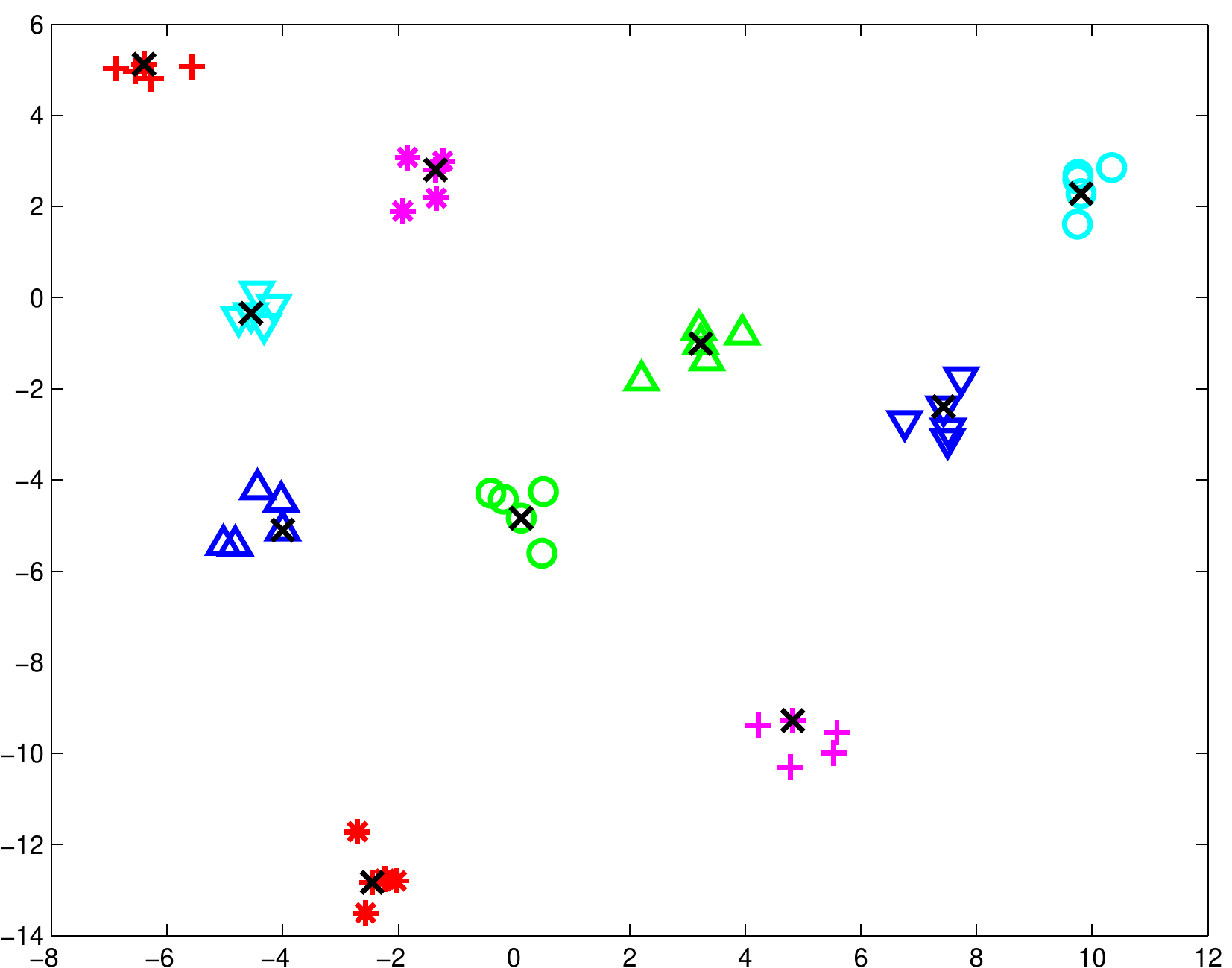}\label{fig:TenClouds_M_GL_lambda20_var2_s2}}
\caption{Output clustering via the solution of the sum--of--norms relaxation \eqref{probl:OMT_clustering_matr_no_p1_relax} for different values of the parameter 
$\lambda$. 
The clusters representatives are denoted by a black cross. Different shapes and colors have been used to denote points belonging to different clusters. }
\label{fig:GL_different_lambda_10clusters}
\end{figure*}

\begin{figure*}[htp]
\centering
\subfloat[][$\lambda=1500$]{\includegraphics[width=0.30\textwidth]{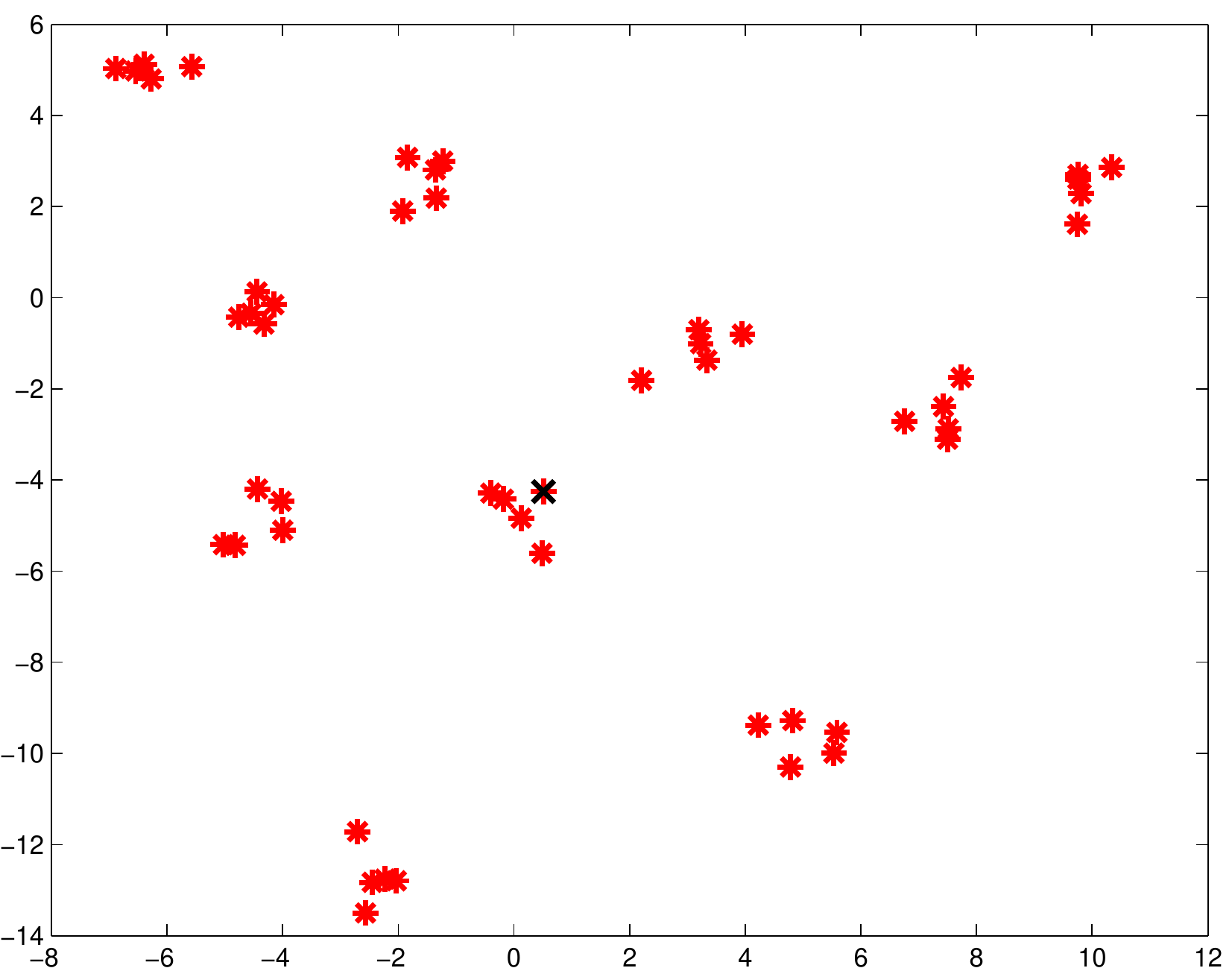}\label{fig:TenClouds_M_LP_lambda500_var2_s2}}\hspace{2mm}
\subfloat[][$\lambda=300$]{\includegraphics[width=0.30\textwidth]{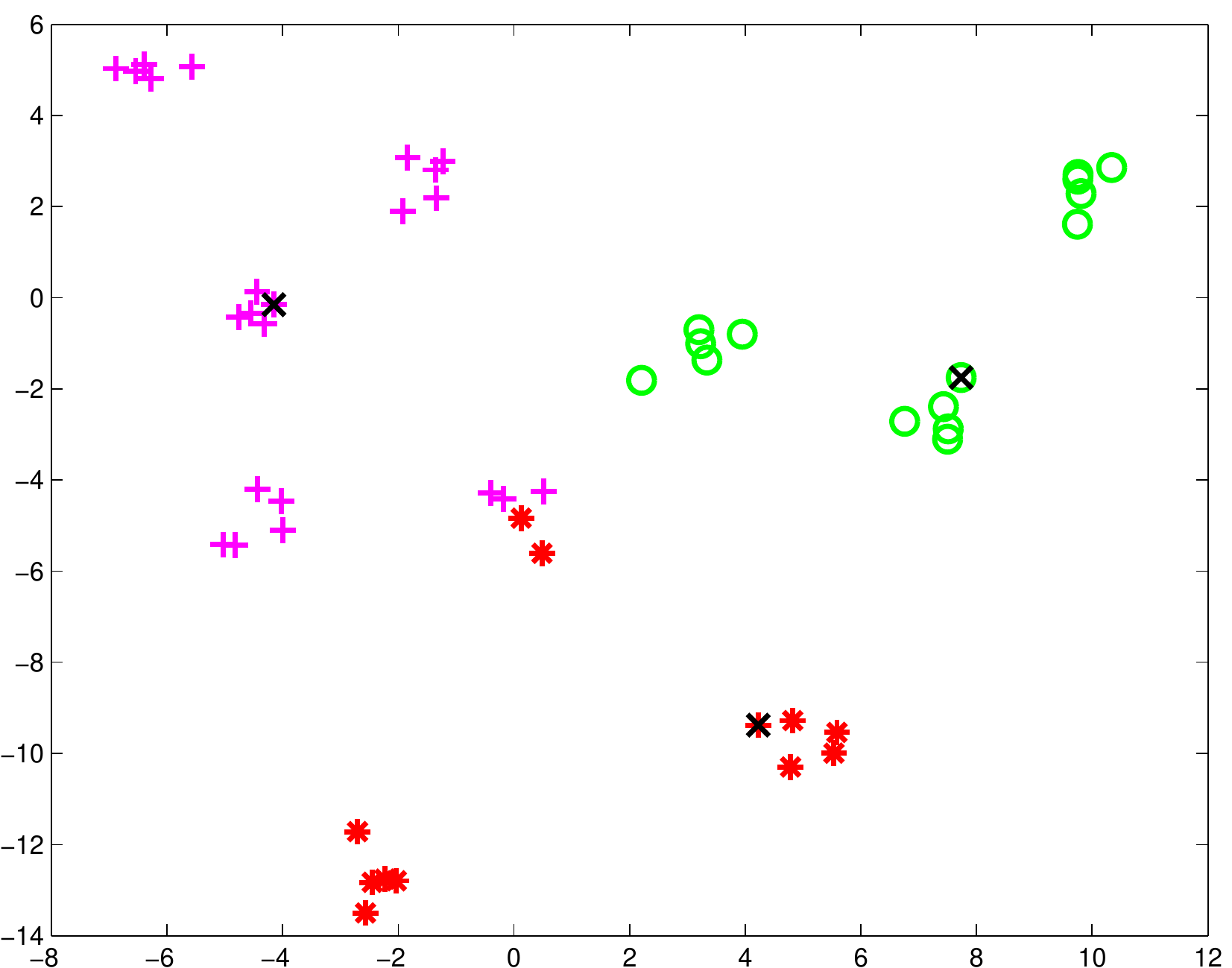}\label{fig:TenClouds_M_LP_lambda100_var2_s2}}
\\\subfloat[][$\lambda=30$]{\includegraphics[width=0.30\textwidth]{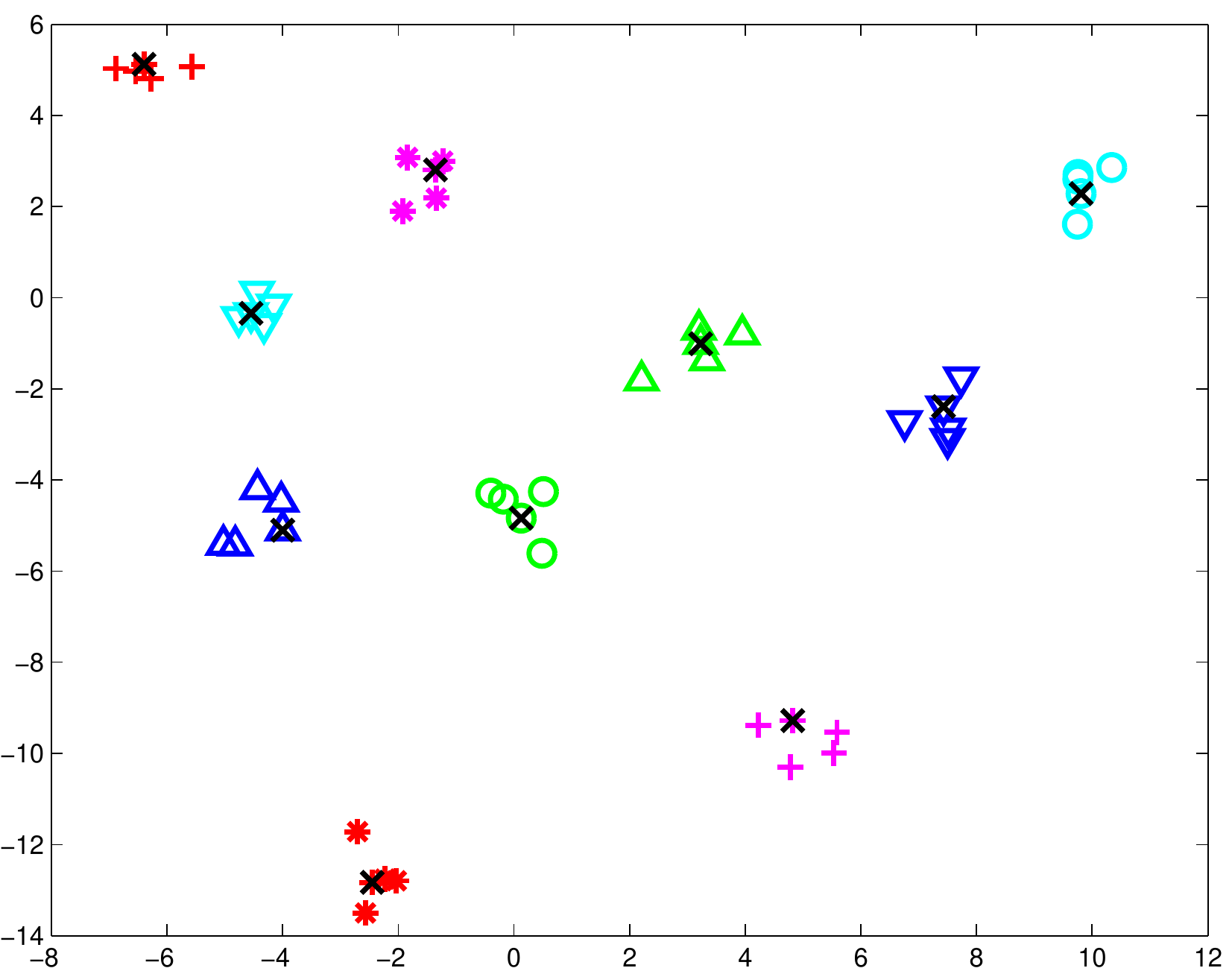}\label{fig:TenClouds_M_LP_lambda50_var2_s2}}\hspace{2mm}
\subfloat[][$\lambda=10$]{\includegraphics[width=0.30\textwidth]{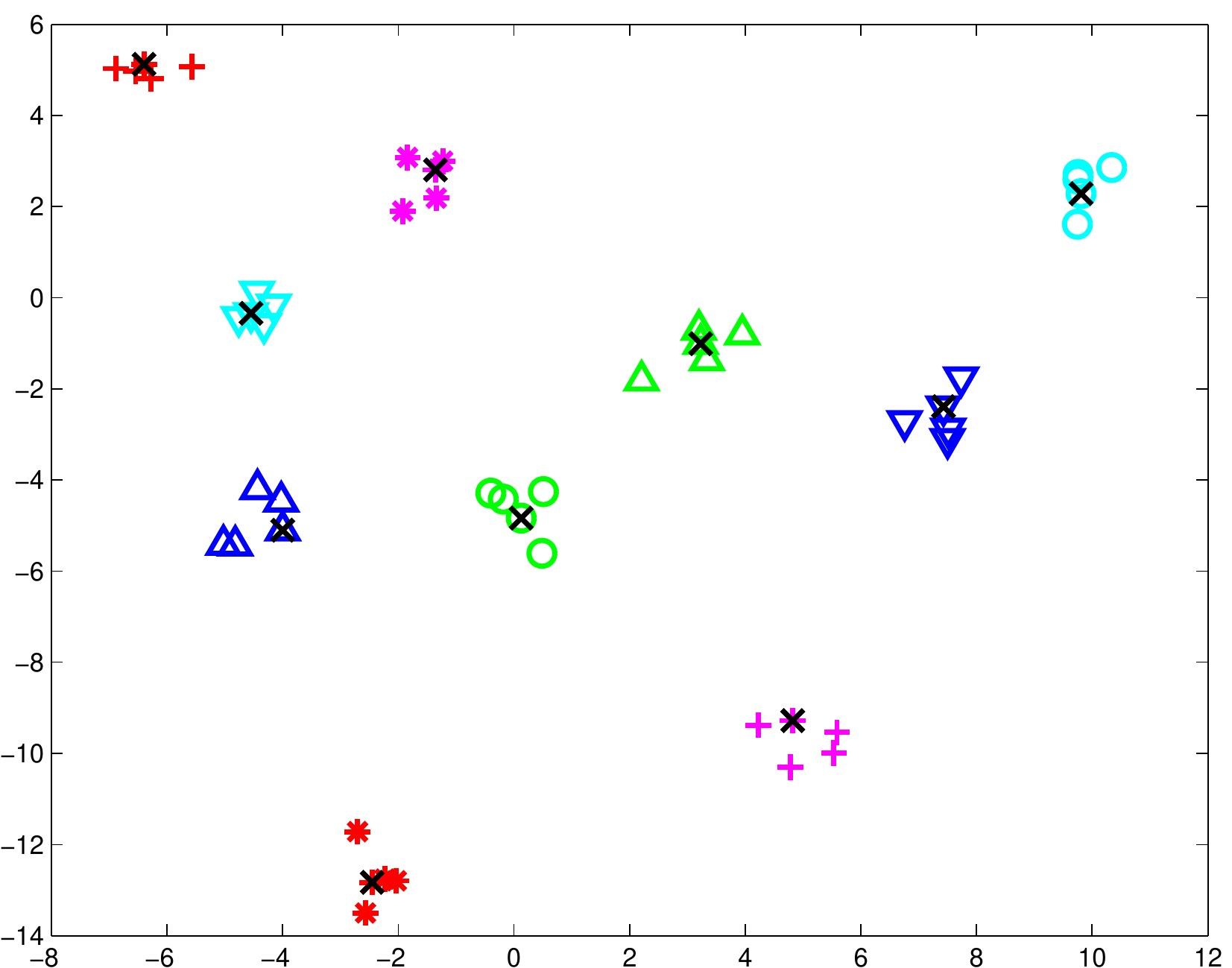}\label{fig:TenClouds_M_LP_lambda20_var2_s2}}
\caption{Output clustering via the solution of the LP formulation \eqref{prob:y_relax} for different values of the parameter $\lambda$.  
The clusters representatives are denoted by a black cross. Different shapes and colors have been used to denote points belonging to different clusters. }
\label{fig:LP_different_lambda_10clusters}
\end{figure*}

\begin{figure*}[htp]
\centering
\subfloat[][$\lambda=150$]{\includegraphics[width=0.30\textwidth]{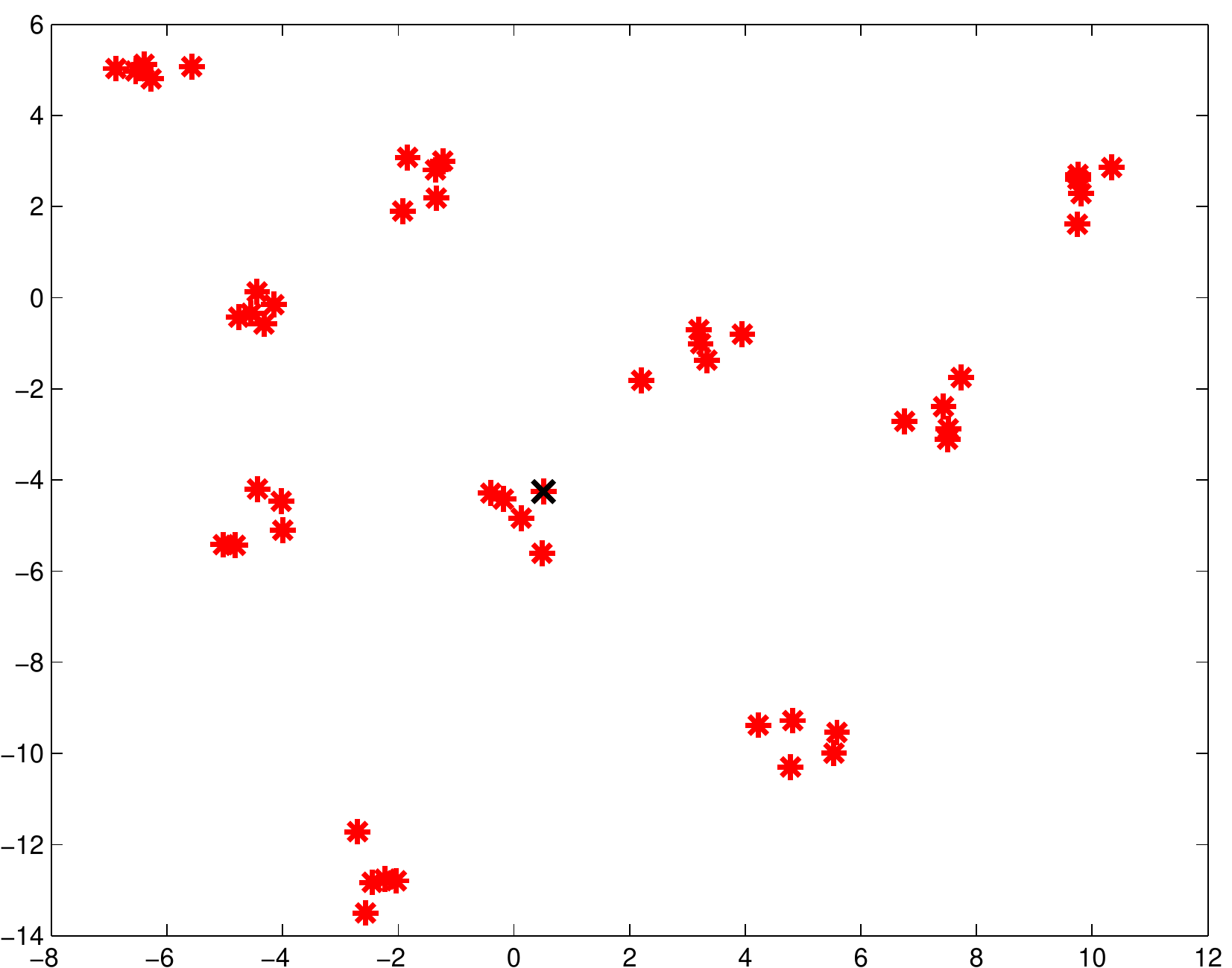}\label{fig:TenClouds_M_inf_lambda150_var2_s2}}\hspace{2mm}
\subfloat[][$\lambda=30$]{\includegraphics[width=0.30\textwidth]{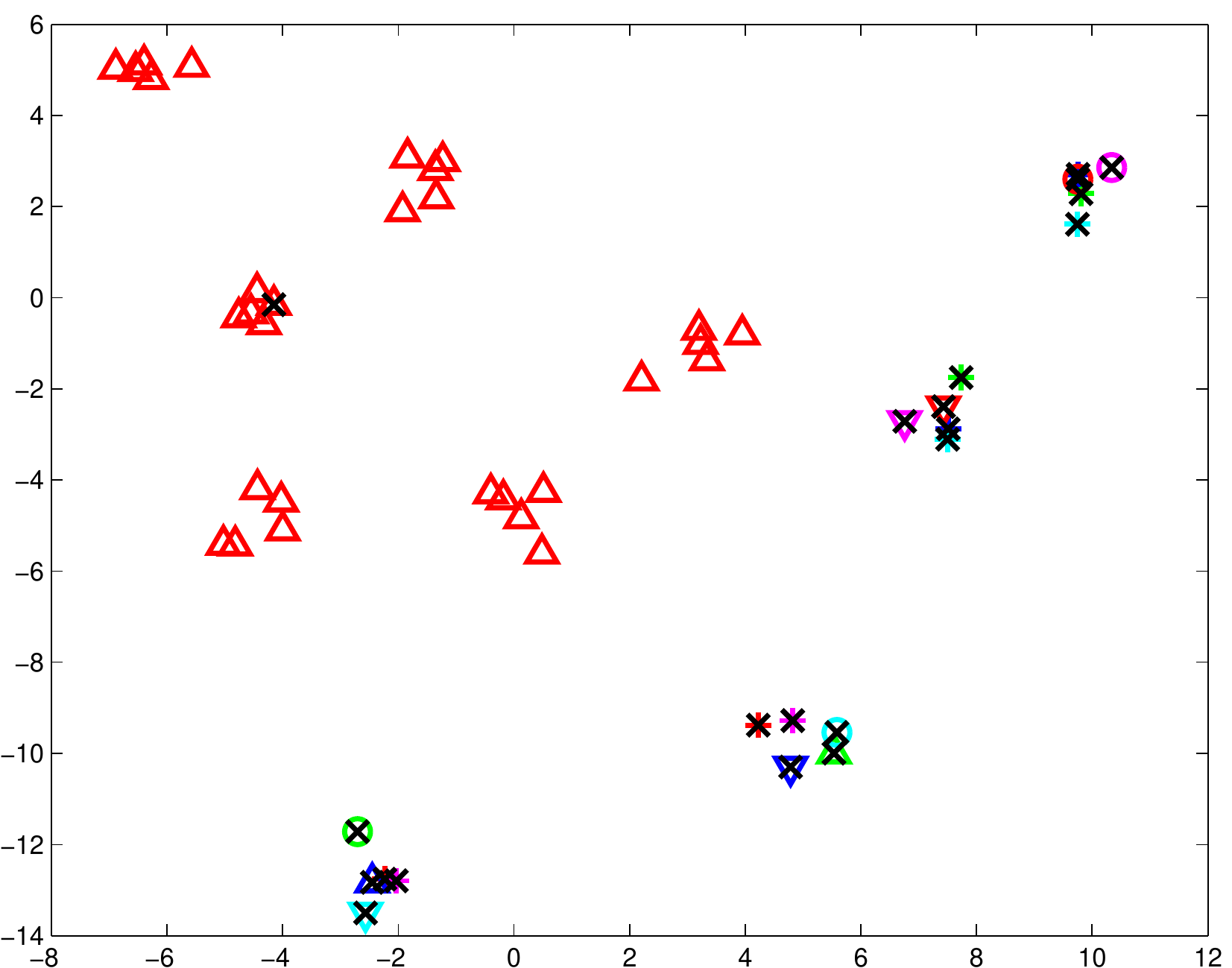}\label{fig:TenClouds_M_inf_lambda30_var2_s2}}
\\\subfloat[][$\lambda=8$]{\includegraphics[width=0.30\textwidth]{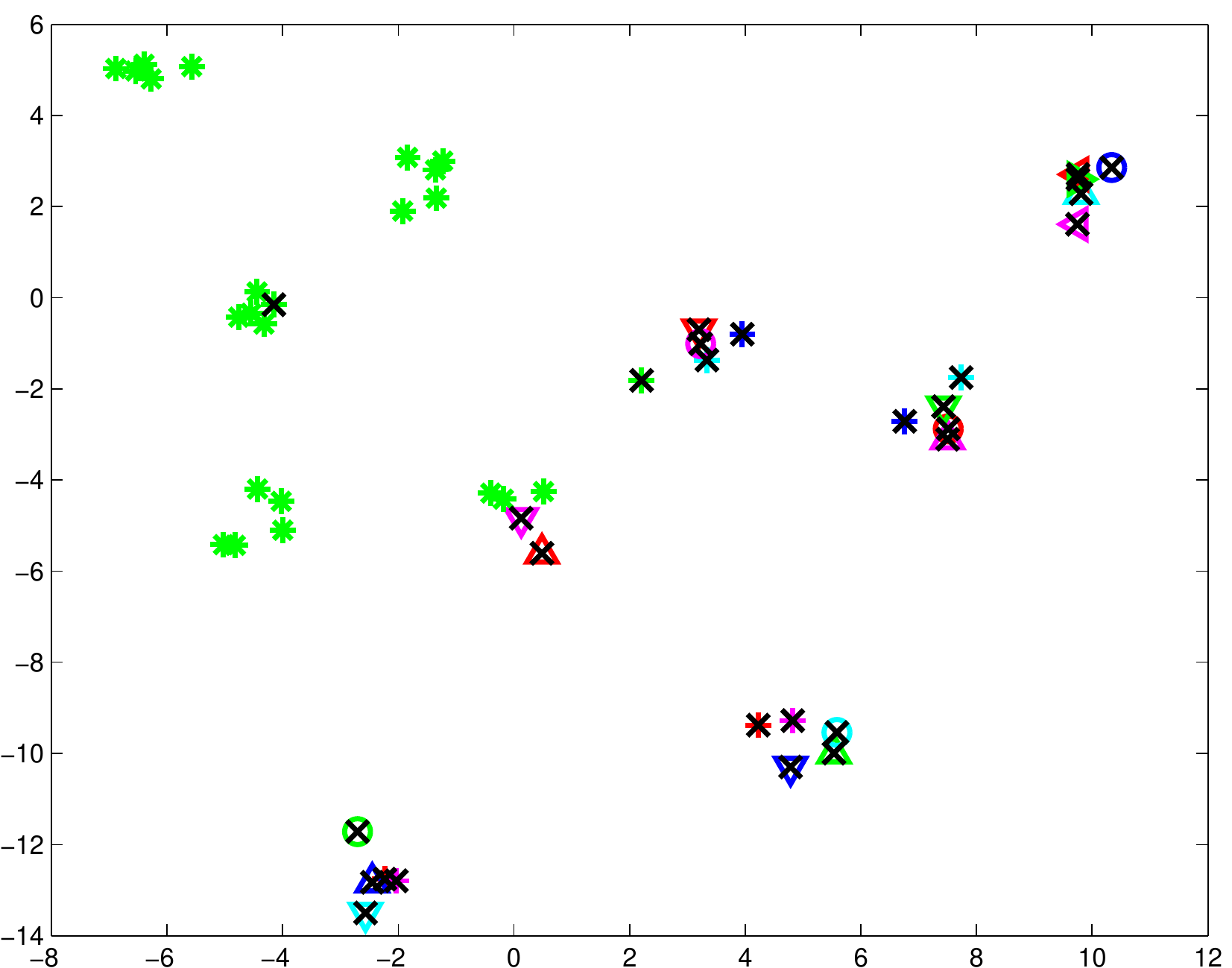}\label{fig:TenClouds_M_inf_lambda8_var2_s2}}\hspace{2mm}
\subfloat[][$\lambda=2$]{\includegraphics[width=0.30\textwidth]{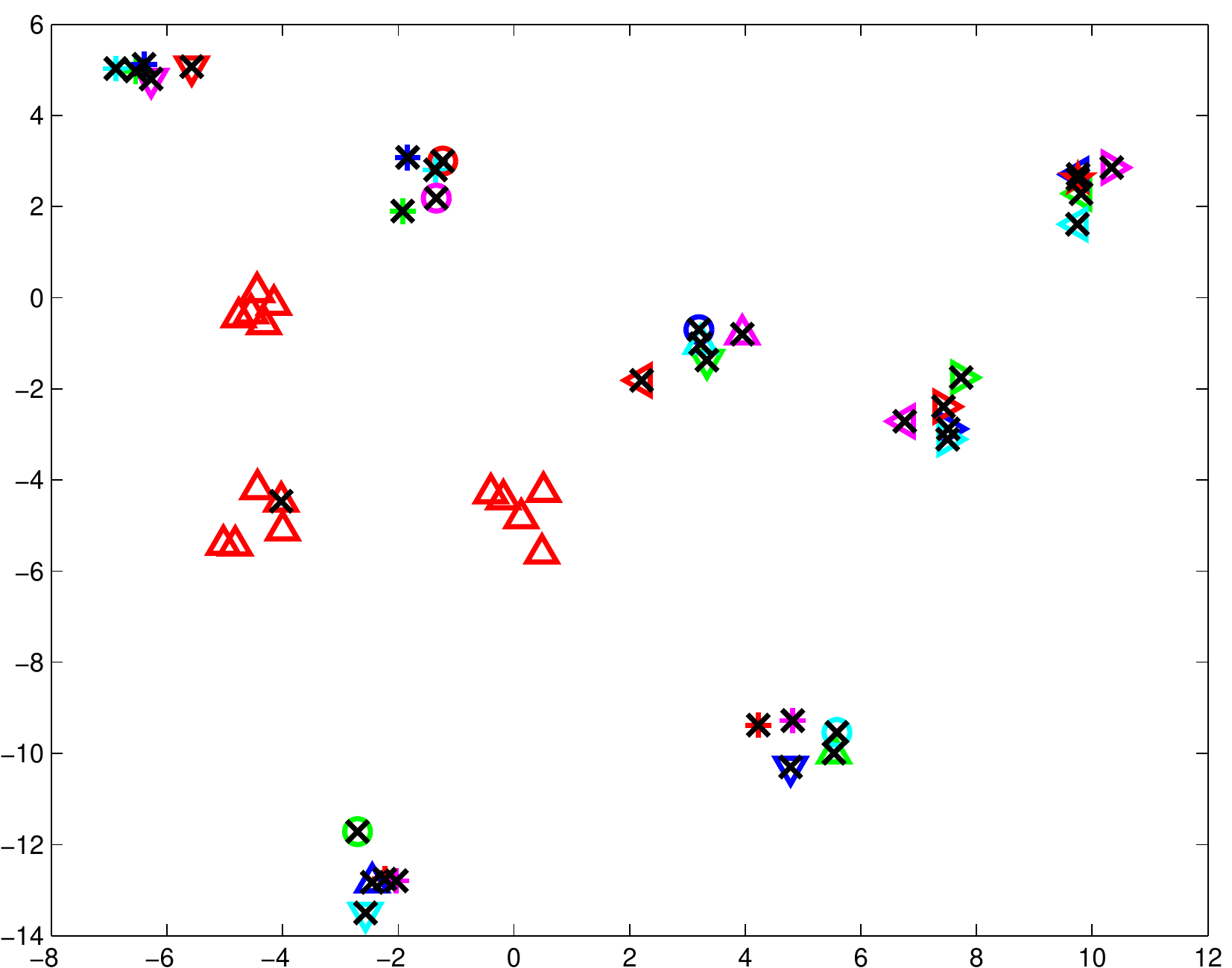}\label{fig:TenClouds_M_inf_lambda2_var2_s2}}
\caption{Output clustering via the solution of the $\ell_\infty$--norm relaxation \eqref{probl:OMT_clustering_matr_ElGhaoui_convex}  for different values of the parameter $\lambda$.  
The clusters representatives are denoted by a black cross. Different shapes and colors have been used to denote points belonging to different clusters. }
\label{fig:inf_different_lambda_10clusters}
\end{figure*}

The three relaxations have also been applied to a data set with 10 clusters generated by sampling 10 Gaussian distributions with means 
\begin{align*}
\big\{
&\left(-2.5 ,\, -12.5\right),
\left(5 ,\, -10\right),
\left(0  ,\, -5\right),
\left(-4.5 ,\, -5\right),
\left(-5  ,\,    0\right),\\
&\left(-6 ,\, 5\right),
\left(-1.5 ,\,  2.5\right),
\left(3.5 ,\, -1\right),
\left(7.5 ,\,-2.5\right),
\left(10 ,\,  2.5\right)\big\}
\end{align*} 
and common covariance 
$$
\bmat 0.2 & 0 \\ 0& 0.2\emat \,.
$$
The experimental results are reported in Figure \ref{fig:GL_different_lambda_10clusters}, \ref{fig:LP_different_lambda_10clusters} 
and \ref{fig:inf_different_lambda_10clusters}, respectively. 
Once again, the sum--of--norms relaxation and the relaxation based on integer programming are able to achieve the correct clustering, 
while there are no values of $\lambda$ for which the penalty based on the $\ell_\infty$--norm is able 
to correctly partition the data set in ten clusters.

\section{Conclusions}\label{sec:conclusions}

In this paper, we considered the problem of approximating distributions within the framework of optimal mass transport.  
We focused on approximating sample distributions with distributions having sparse support. 
Standard $\ell_1$ regularization cannot be used to promote sparsity on a probability simplex since the $\ell_1$ norm of a probability measure is always one. 
We proposed relaxations of the cardinality penalty which are applicable to probability simplices.
One of these relaxation has the property to be the tightest convex lower approximation of the original cardinality penalty. 
When applied to a clustering problem, the proposed framework leads to \emph{convex} clustering schemes, 
thus overcoming sensitivity to initialization of classical clustering algorithms such as k--means.  
The proposed relaxations have been tested on a clustering problem with synthetic generated data. 
Both the relaxation based on the sum--of--norms and on the integer programming formulation act effectively in promoting sparsity 
of the clusters representatives distribution and are able to produce the correct clustering. 

\balance

%
%
%


\bibliographystyle{plain}
\bibliography{biblio_OMTclustering}

\begin{thebibliography}{10}

\bibitem{BoydVand2004}
S.~Boyd and L.~Vandenberghe.
\newblock {\em Convex optimization}.
\newblock Cambridge university press, 2004.

\bibitem{Bruckstein2009}
A.M. Bruckstein, D.L. Donoho, and M.~Elad.
\newblock From sparse solutions of systems of equations to sparse modeling of
  signals and images.
\newblock {\em SIAM review}, 51(1):34--81, 2009.

\bibitem{Candes2005}
E.J. Candes and T.~Tao.
\newblock Decoding by linear programming.
\newblock {\em IEEE Transactions on Information Theory}, 51(12):4203--4215,
  2005.

\bibitem{Chandrasekaran2010}
V.~Chandrasekaran, B.~Recht, P.A. Parrilo, and A.S. Willsky.
\newblock The convex algebraic geometry of linear inverse problems.
\newblock In {\em 48th Annual Allerton Conference on Communication, Control,
  and Computing (Allerton), 2010}, pages 699--703, 2010.

\bibitem{CharikarTardosShmoys1999}
M.~Charikar, S.~Guha, {\'E}.~Tardos, and D.B. Shmoys.
\newblock A constant-factor approximation algorithm for the k-median problem.
\newblock In {\em Proceedings of the thirty-first annual ACM symposium on
  Theory of computing}, pages 1--10. ACM, 1999.

\bibitem{CharikarTardosShmoys2002}
M.~Charikar, S.~Guha, {\'E}.~Tardos, and D.B. Shmoys.
\newblock A constant-factor approximation algorithm for the k-median problem.
\newblock {\em Journal of Computer and System Sciences}, 65(1):129--149, 2002.

\bibitem{Chen1998}
S.S. Chen, D.L. Donoho, and M.A. Saunders.
\newblock Atomic decomposition by basis pursuit.
\newblock {\em SIAM journal on scientific computing}, 20(1):33--61, 1998.

\bibitem{Daskin1995}
M.S. Daskin.
\newblock {\em Network and discrete location: models, algorithms, and
  applications.}
\newblock Wiley, 1995.

\bibitem{DreznerHamacher2001}
Z.~Drezner and H.W. Hamacher.
\newblock {\em Facility location: applications and theory}.
\newblock Springer-Verlag, 2001.

\bibitem{Fazel2001}
M.~Fazel, H.~Hindi, and S.P. Boyd.
\newblock A rank minimization heuristic with application to minimum order
  system approximation.
\newblock In {\em Proceedings of the American Control Conference}, volume~6,
  pages 4734--4739, 2001.

\bibitem{JainVazirani2001}
K.~Jain and V.V. Vazirani.
\newblock Approximation algorithms for metric facility location and k-median
  problems using the primal-dual schema and lagrangian relaxation.
\newblock {\em Journal of the ACM (JACM)}, 48(2):274--296, 2001.

\bibitem{Kantorovich1942}
L.V. Kantorovich.
\newblock On the transfer of masses.
\newblock In {\em Dokl. Akad. Nauk. SSSR}, volume~37, pages 227--229, 1942.

\bibitem{Lashkari2007}
Danial Lashkari and Polina Golland.
\newblock Convex clustering with exemplar-based models.
\newblock In {\em Advances in {N}eural {I}nformation {P}rocessing {S}ystems},
  pages 825--832, 2007.

\bibitem{LinVitter1992}
J.H. Lin and J.S. Vitter.
\newblock e-approximations with minimum packing constraint violation.
\newblock In {\em Proceedings of the twenty-fourth annual ACM symposium on
  Theory of computing}, pages 771--782. ACM, 1992.

\bibitem{Monge1781}
G.~Monge.
\newblock {\em M{\'e}moire sur la th{\'e}orie des d{\'e}blais et des remblais}.
\newblock De l'Imprimerie Royale, 1781.

\bibitem{ElGhaoui2012NIPS}
M.~Pilanci, L.~El~Ghaoui, and V.~Chandrasekaran.
\newblock Recovery of sparse probability measures via convex programming.
\newblock In {\em Advances in {N}eural {I}nformation {P}rocessing {S}ystems},
  pages 2429--2437, 2012.

\bibitem{Rachev1998-I}
S.T. Rachev and L.~R{\"u}schendorf.
\newblock {\em Mass Transportation Problems: Volume I: Theory}.
\newblock Springer, 1998.

\bibitem{Rachev1998-II}
S.T. Rachev and L.~R{\"u}schendorf.
\newblock {\em Mass Transportation Problems: Volume II: Applications}.
\newblock Springer, 1998.

\bibitem{Tibshirani1996}
R.~Tibshirani.
\newblock Regression shrinkage and selection via the lasso.
\newblock {\em Journal of the Royal Statistical Society. Series B
  (Methodological)}, pages 267--288, 1996.

\bibitem{Villani2003}
C.~Villani.
\newblock {\em Topics in optimal transportation}, volume~58.
\newblock American Mathematical Society, 2003.

\bibitem{Villani2008}
C.~Villani.
\newblock {\em Optimal transport: old and new}, volume 338.
\newblock Springer, 2008.

\bibitem{YuanLin2005}
M.~Yuan and Y.~Lin.
\newblock Model selection and estimation in regression with grouped variables.
\newblock {\em Journal of the Royal Statistical Society: Series B (Statistical
  Methodology)}, 68(1):49--67, 2005.

\end{thebibliography}

\end{document}